\documentclass[12pt]{article}

\usepackage{graphicx}
\usepackage{epstopdf}
\usepackage[body={17.5cm, 22cm},right=2.2cm]{geometry}
\usepackage{amssymb}
\usepackage{amsmath}

\newcommand\ee{\end{equation}}
\newcommand\be{\begin{equation}}
\newcommand\eea{\end{eqnarray}}
\newcommand\bea{\begin{eqnarray}}

\def\e{{\rm e}}

\def\d{\partial}
\def\l{\left(}
\def\r{\right)}
\def\beq{\begin{equation}}
\def\eeq{\end{equation}}

\def\barr{\begin{array}}
\def\earr{\end{array}}

\begin{document}

\setcounter{page}{0}
\thispagestyle{empty}
\begin{flushright}
CERN-PH-TH/2008-121\\
\end{flushright}

\vspace*{1cm}

\begin{center}
{\Large
\bf{
Superluminal Travel Made Possible\\
[0.3cm]
(in two dimensions)
}}\\[1cm]
{\large Sergei Dubovsky$^{\rm a, c}$, Sergey Sibiryakov$^{\rm b,c}$
}
\\[0.5cm]

{\small \textit{$^{\rm a}$ Jefferson Physical Laboratory, \\ Harvard University, Cambridge, MA 02138, USA}}

\vspace{.2cm}
{\small \textit{$^{\rm b}$ CERN Theory Division, CH-1211 Geneva 23, Switzerland}}

\vspace{.2cm}
{\small \textit{$^{\rm c}$ Institute for Nuclear Research of the Russian Academy of Sciences, \\
        60th October Anniversary Prospect, 7a, 117312 Moscow, Russia}}
\end{center}
\vspace{0.5cm}

\begin{abstract}

We argue that superluminal signal propagation
is possible in consistent Poincare invariant quantum field theories in
two space-time 
dimensions, provided spatial parity is broken. This happens due to
existence of the 
``instantaneous'' causal structure, with one of the
light cone variables being a global time. In two dimensions this
causal structure is invariant under the Poincare group if one gives
up the spatial parity. As a non-trivial example of a consistent
interacting quantum field theory with this causal structure we discuss
a non-linear $SO(1,1)$ sigma-model, where $SO(1,1)$ is the Lorentz
symmetry. We show that this
theory is asymptotically free and argue that this
model is also well defined non-perturbatively, at least for some
values of parameters. It provides an example
of a microscopic Poincare invariant quantum field theory with local
action, but non-local physical properties.  Being coupled to gravity
this ``instantaneous'' theory mixes with the Liouville field. If
proves to be consistent, the resulting model can be used to construct 
(non-critical) string theories with very unconventional
properties by introducing the instantaneous causal structure on the
world-sheet.

\end{abstract}

\vfill

\setcounter{footnote}{0}
\newpage
\section{Introduction}
\label{sec:intro}
The question we will address in the current paper is
\begin{center} {\it Whether a superluminal signal propagation is
    possible in a consistent Poincare invariant quantum field theory?
  }
\end{center}
This question may appear rather provocative as it seems to be known
that the answer is {\it No}. Also the
physics to which this question will lead us 
turns out to be quite unconventional, so
let us start with explaining our motivations. The very first
motivation (or, rather, an excuse) for asking this question is that we
claim that the answer is {\it Yes}: superluminal, and even 
instantaneous signal propagation is possible, at least in two
space-time dimensions with broken spatial parity.

Apart from this rather unexpected answer there is a number of other
motivations more directly linked to the known physics. First, there
are reasons to suspect that locality is only an approximate notion in
gravitational theories. Probably the strongest arguments supporting
this viewpoint are the absence of sharply defined local observables in
theories with dynamical gravity, and the indication that the information recovery
during a black hole evaporation requires a certain degree of
non-locality (see e.g., \cite{Giddings:2007ie,ArkaniHamed:2007ky} for
recent discussions). On the other hand, in a Poincare invariant theory
non-locality in space would imply also non-locality in time that
appears to  be at odds with unitary time evolution and
causality. To analyze this problem it is useful to construct explicit models with non-local
properties. In this paper we will present explicit models exhibiting instantaneous signal propagation, that definitely qualifies as a
non-local effect. 
To the best
of our knowledge, these models provide the first example of Poincare invariant microscopic
quantum field theories with a truly non-local behavior, if as yet in two
dimensions.

A more general motivation to ask a question like that
is a theoretical curiosity triggered by the following
considerations. In the early years of GUT and string theory there was
a
hope that the consistency conditions for 
coexistence of gravity and quantum field theory
uniquely fix the Standard Model as the only possible low energy
effective field theory in a world with gravity. 
However, later developments such as D-branes~\cite{Polchinski:1995mt}
revealed that the power of mathematical consistency was overestimated
and a number of low energy field theories that can be embedded in a
microscopic gravitational theory is much larger.  This trend has
apparently reached its top with the discovery of a vast landscape of
metastable de Sitter vacua~\cite{Shamit}. 

At this point it is natural to ask where is the boundary (if any)
of the set of consistent gravitational theories. 
A plausible answer to this question may be the
``swampland'' idea~\cite{Vafa:2005ui}. Namely, it well may be that
instead of directly predicting the spectrum and/or coupling constants
of the Standard Model, self-consistency considerations imply a set of
relations between them, invisible at the level of an effective field
theory. Effective field
theories where these relations do not hold would belong to the
``swampland'' and cannot be UV completed to a theory of quantum
gravity. Along these lines several examples of surprising limitations
on effective field theories were found
\cite{ArkaniHamed:2006dz,Ooguri:2006in,Dvali:2008tq}.

Still, an adventurous person may challenge this point of view and
suggest that deep inside the swampland one may discover new islands of
theories, that are totally different from all what we know at the
moment (``wonderland''). In other words one may wonder whether 
there exist UV complete theories
where
the principles of quantum field theory we usually take for granted are
violated? Clearly, the impossibility of a superluminal signal
propagation is one of such principles.

In fact, a number of effective field theories were constructed
recently that will either turn out to provide quite interesting
examples of the swampland inhabitants, or may be a hint that
gravitational theories with rather exotic properties exist. These
theories resulted from attempts to find a consistent modification of
gravity at long distances, to large extent motivated by the
cosmological constant problem. They include the brane world DGP model
\cite{Dvali:2000hr} as well as four dimensional models describing
gravity in the Higgs phase, such as the ghost
condensate~\cite{ArkaniHamed:2003uy} and more general models of
massive gravity \cite{Dubovsky:2004sg} (see \cite{Rubakov:2008nh} for
a review). A closely related class of models is the Einstein-aether
theory~\cite{Jacobson:2000xp} and its generalizations.

These models appear consistent from the effective field theory
viewpoint and provide a number of interesting phenomenological
signatures (such as the anomalous precession of the Moon perihelion in
the DGP model~\cite{Dvali:2002vf,Lue:2002sw} or a strong monochromatic
gravitational wave signal due to primordial massive gravitons~\cite{Dubovsky:2004ud}).  Also they provide a number of
interesting theoretical possibilities, for instance, opening a room
for an alternative to inflation~\cite{Creminelli:2006xe}. On the other
hand, it is not clear at all whether these models can be UV completed
in a consistent microscopic quantum theory. In fact, there are good
reasons~\cite{Adams:2006sv,Dubovsky:2006vk,Eling:2007qd}, based on
considerations related to locality, causality and black hole physics,
why such a UV completion is extremely hard to achieve.  To find a way
around these problems (or to conclusively demonstrate that such a way
doesn't exist) requires to further scrutinize the standard lore about causality in
Lorentz invariant quantum field theory.

Clearly, it is straightforward to construct consistent quantum
theories with faster-than-light or even instantaneous signal
propagation if one gives up Lorentz invariance. Most non-relativistic
quantum mechanical systems have this property.  The real challenge is to
incorporate superluminal propagation in a theory with gravity. To
stress the importance of existence of the universal causal cone in
gravitational theory, it suffices to note that if superluminal
propagation were allowed, information could be extracted from black
holes already at the classical level  \cite{Dubovsky:2006vk,Babichev:2006vx,Dubovsky:2007zi}
and a remarkable success of 
black hole thermodynamics in general relativity and string theory
would appear to be an unnecessary coincidence. Indeed, the
conventional black hole thermodynamics breaks down if there is no
causal cone, universal for all
fields~\cite{Dubovsky:2006vk,Eling:2007qd}.

If one starts with a gravitational theory and decouples gravity by
sending $M_{Pl}$ to infinity, one ends up with a Poincare invariant
quantum field theory. That's why we will focus on Poincare invariant
field theories and study whether superluminal propagation is
compatible with the Poincare invariance. Well-known arguments reviewed in
Sec.~\ref{sec:generalities} suggest that it is impossible.
However, there is a couple of loopholes in these
arguments leaving a chance for superluminal travel.

The first loophole is that in two space-time dimensions, if one gives
up the
spatial parity, there are two different causal structures
compatible with the remaining (connected) part of the Poincare
group. The first is the standard one, and the second is the
``instantaneous'' causal ordering, such that for a given point the
half-plane with a larger value of the global light cone time $x^+$ is
in the causal future, and the half-plane with a smaller value of $x^+$
is in the causal past. We will refer to this structure as $x^+$-ordered causal structure.
A natural question arises whether consistent
quantum field theories with the second choice of the causal structure
exist. We will answer affirmatively to this question and hope to
convince the reader that these theories may be quite
interesting. This is the main result of the present paper.

Another loophole arises due to the possibility of spontaneous breaking
of Lorentz invariance. Models describing gravity in the Higgs phase
 realize exactly this idea.  As we said, in spite of a significant
progress achieved recently in constructing such models at the
effective field theory level, none of them has been embedded into a UV
complete microscopic theory so far.
It appears easier to achieve this goal in two space-time dimensions
due to a very simple reason: many field theories that are
non-renormalizable and require UV completion in higher dimensions can
be renormalized at $D=2$. One may think that this hope is
doomed due to the Coleman--Mermin--Wagner (CMW) theorem
\cite{Mermin:1966fe,Coleman:1973ci} establishing that spontaneous
breaking of a continuous symmetry is not possible in two dimensions.
However, as we argue below, in spite of the restoration of the Lorentz
invariance at long distances, the short
distance physics captures all relevant and interesting properties
following from the spontaneous symmetry breaking.

Thus two somewhat different reasons --- special geometrical properties
and better chances to construct a UV complete theory with spontaneous
Lorentz breaking --- lead us to study a possibility of superluminal
travel in two space-time dimensions. 
We start by discussing in Sec.~\ref{sec:generalities} how the standard
arguments against superluminal signal propagation are avoided in
2d. In Sec.~\ref{sec:aether} we introduce the model which we consider
in the rest of the paper. This is the 
so-called Einstein-aether theory, a model describing a dynamical vector
field that develops a Lorentz-violating vev. This model has been recently
studied in two dimensions at the classical level in
\cite{Eling:2006xg}, while our main interest here is whether it can be
promoted to a consistent quantum theory. We identify four different
regimes of this model corresponding to different choices of
parameters. Two choices explicitly break the spatial parity and
naturally enjoy the causal structure based on time-ordering with
respect to the light-cone coordinate $x^+$, while in the third and the fourth cases the action is
parity symmetric.

We show in
Sec.~\ref{sec:oneloop} that at the perturbative level all four
parameter choices give rise to renormalizable asymptotically free
theories. However, beyond the perturbation theory the four cases are
quite different. We consider them separately in
Sec.~\ref{sec:brokenparity}. 

We start with the case where violation of parity is, in a sense,
maximal. This turns out to be the cleanest example of a consistent
theory with the
instantaneous causal structure.  This theory allows a (non-standard)
Wick rotation resulting in a path integral with a positive definite
real part of the Euclidean action.  Under the renormalization group (RG) 
it flows from a free
theory with the standard causal structure in the UV to a weakly coupled
instantaneous theory in the IR. 
Physically,
this implies that an observer in such a theory by performing short
scale measurements can be tricked into thinking that she lives in a
world with the standard causal structure. However, after some time (or
increasing the precision of the measurements) she will find out
apparent causal paradoxes that are removed after realization what the
actual causal structure is. Given that the true causal structure is
instantaneous, this model provides an example of a genuinely non-local
Poincare invariant quantum field theory.

Then we turn to the other class of parity breaking Einstein-aether
models. In the minimal version they exhibit the conventional causal structure, but 
by introducing an additional field can be easily turned into theories obeying the $x^+$-ordered causal structure.
The
interesting property of these theories is that at the classical level
they are invariant under the right moving part of the Virasoro algebra
and under the global part of the left mover's Virasoro algebra. At
the quantum level, in spite of being derivatively coupled, they can be
renormalized by normal ordering. This gives rise to a hope that one
may find solvable models with the instantaneous causal structure.

For completeness, we discuss also the parity preserving cases,
though they do not provide theories obeying the $x^+$-ordered causal
structure. 
For one choice of parameters the causal structure is the standard one. 
In IR the theory flows
to strong coupling which suggests that a mass gap is
generated. Perturbatively, the theory is stable. However, we find that
the classical Hamiltonian is unbounded from below indicating
a non-perturbative instability of the semiclassical vacuum. This may or
may not imply inconsistency of the model for this parameter choice.

Finally, we consider the second parity preserving choice of parameters that leads to a theory where a fixed causal structure is absent altogether. This case is the most subtle of all. As we discuss, it has much in common with the gravitational theories. In particular, the causal structure itself is dynamical: it is determined by a background under consideration. Though the action in this case is parity symmetric, at the quantum level the theory can be consistent if the spatial parity is spontaneously broken. This may be regarded as the two-dimensional analog of a spontaneous breaking of Lorentz symmetry in higher dimensions.

In Sec.~\ref{sec:gravity} we outline what happens when the above
models 
are coupled to gravity. We show that in the conformal gauge the
Einstein-aether sector produces a contribution to the Liouville action
of the dilaton equivalent to that of a single two-dimensional scalar
boson. Besides, the 
Einstein-aether
mode mixes with the Liouville mode.
It remains to be understood whether the resulting model allows for a
consistent quantization.

We conclude in Sec.~\ref{sec:conclusions} by summarizing 
the lessons obtained from our study, and point out what we think are the promising directions
for future work.

\section{Superluminality in two dimensions}
\label{sec:generalities}
\subsection{Why superluminal travel is hard in general}
\label{subsec:generalities}
\begin{figure}[t]
\centering
\includegraphics[viewport=-0 245 800 520,width=1.0\textwidth,clip]{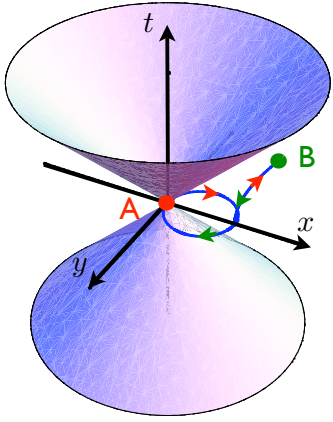}
\caption{\small\it Any two spacelike separated points  A and B can be interchanged by a combination of a Lorentz boost and
 a spatial rotation. Consequently, no Poincare invariant causal ordering exists such that A and B are causally connected.}
\label{noway}
\end{figure}

Let us start with a brief review of well-known arguments why
superluminal travel is in general problematic in Poincare-invariant
theories (the bulk of this discussion follows
\cite{Adams:2006sv}). Let us first assume that the Lorentz invariance is
unbroken and it is possible to transfer a signal from a space-time
point A to another point B at the space-like
separation. This possibility implies that the point A is in the causal
past of the point B.  However, there always exists a Lorentz
transformation that exchanges points A and B with each other (see Fig.~\ref{noway}) so that
one concludes that the point B should also be in the causal past of
the point A.  This clearly does not make sense or, formally, in this
case it is impossible to introduce a causal ordering of events. This
proves that superluminal travel is incompatible with unbroken Lorentz
invariance.

One proposal to get around this problem would be to say that
superluminal travel is not possible in the Lorentz-invariant vacuum,
but excitations in non-trivial backgrounds can nevertheless propagate
outside the Minkowski light-cone. A simple example of a classical
Lorentz-invariant theory with this type of behavior is provided by the
following action describing a single scalar field $\phi$, 
\be
\label{Xaction}
S=\int d^Dx \;\Lambda^4P(X)\;, 
\ee 
where\footnote{We use the metric signature $(+,-,-,\ldots)$.} 
$X=(\d_\mu\phi)^2$. At the
quantum level for non-trivial functions $P$ the theory (\ref{Xaction})
is non-renormalizable and should be considered as an effective low
energy theory with a cutoff $\leq \Lambda$.  In this model scalar
perturbations $\pi$ propagate with the speed of light in the
Lorentz-invariant vacuum $\phi=0$; in a non-trivial background
their propagation is described by the wave equation, \be
\label{scaleq}
\Box_G\pi=0\;,
\ee 
where $\Box_G$ is the d'Alambertian with respect to the effective
``acoustic'' metric (cf. \cite{Babichev:2007dw})
\be
\label{effective_metric}
G_{\mu\nu}=\Omega\left(\eta^{\mu\nu}
+\frac{2P''}{P'}\d^\mu\phi\d^\nu\phi\right)^{-1}=
\Omega\l\eta_{\mu\nu}-\frac{2P'' \d_\mu\phi\d_\nu\phi}{P'+2P''X}\r\;,
\ee
with the conformal factor given by
\[
\Omega=\big(P'\l P'+2{P'' X}\r\big)^{1/D-2}\;.
\]
In order to avoid unacceptable ghost or gradient instabilities one needs 
\be
\label{goodness_condition}
P'>0\;,\;\;P'+2P''X>0\;.
\ee
However, there is no obvious pathology if at some value $X_0$ the
second derivative of $P$ turns negative, $P''(X_0)<0$. In the
corresponding backgrounds (for instance, one can take
$\phi=X_0^{1/2}t$)
the causal cone of the
effective metric (\ref{effective_metric}) is broader than the
Minkowski light cone and scalar perturbations can propagate faster
than light.
 
 Note that this kind of field theories have a peculiar property that the space-time events  don't possess any fixed causal ordering independent of the field configuration. This property is definitely very unusual in the field theory context, however, we are used to this situation in gravitational theories.  In particular, at the classical level the absence of the fixed causal structure doesn't prevent one from formulating an initial value problem, provided  one uses surfaces that are space-like with respect to the acoustic metric (\ref{effective_metric}) as a set of Cauchy slices (see \cite{Babichev:2007dw} for a recent discussion
).
 
A problem with this proposal is that while it appears
viable at the level of effective field theory, it is very difficult to
imagine a UV completion of the corresponding 
effective field theories. Clearly, effective field theories of this kind cannot
arise as the low energy limit of conventional Poincare-invariant 
renormalizable field
theories, where microcausality (and, as a consequence, subluminal
propagation in all backgrounds) is built in by
construction. Moreover, if arbitrary small deformation of the Lorentz
invariant vacuum is enough to reach a background with superluminal
propagation (this is the case if $P''(0)<0$), then the $2\to 2$
scattering amplitude in the model (\ref{Xaction}) does not enjoy the
conventional analytical properties, that hold not only in any
renormalizable field theory, but in the string theory (at least at
weak coupling) as well \cite{Adams:2006sv}.

\begin{figure}[t]
\centering
\includegraphics[width=0.7\textwidth,clip]{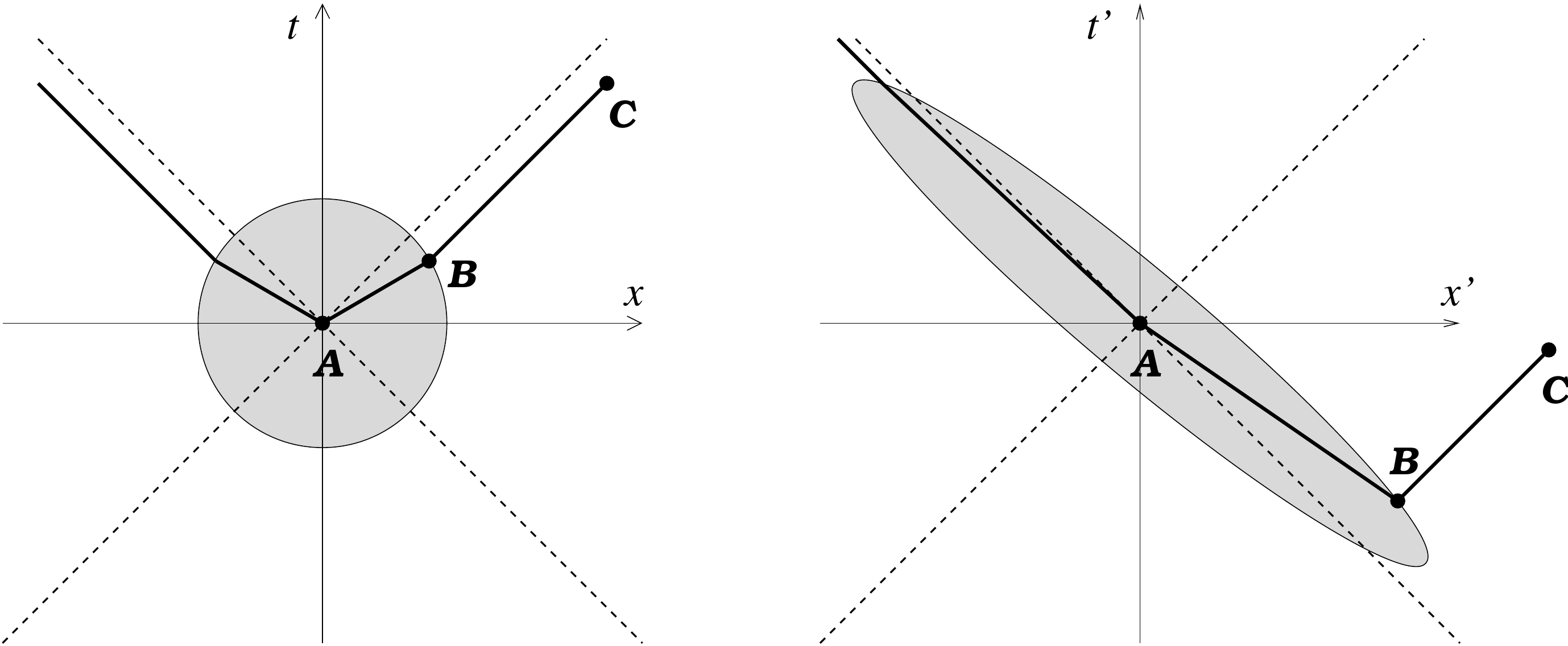}
\put(-310,0){a)}
\put(-130,0){b)}
\caption{ \small\it {\rm a)} A non-trivial field configuration gives rise to a  finite space-time region with superluminal propagation.
{\rm b)} A large boost of such a configuration results in the signal propagation backward in time.
}
\label{broader}
\end{figure}

There is yet another problem with superluminal
travel in this model. Namely, it is possible to find configurations of
the scalar field within the regime of validity of the effective field theory,
such that the effective metric (\ref{effective_metric}) exhibits
closed time-like curves, so that the initial value problem ceases to be globally well-defined.
The idea of this construction is similar to the idea of the paradox
demonstrating impossibility of superluminal travel in the Lorentz
invariant vacuum. One starts with creating a finite region of space
where the causal cone is broader than the Minkowski light cone
(Fig.~\ref{broader}a). If one now performs a large enough boost of this
field configuration, scalar perturbations will propagate back with
respect to the Minkowski time inside the superluminal region
(Fig.~\ref{broader}b). This is not yet a problem by itself, a
non-trivial scalar field configuration spontaneously breaks Lorentz
invariance and the causal structure induced by the effective metric
(\ref{effective_metric}) should be used to define the causal ordering in
such a background.  In more physical terms, one should use the fastest
available particles --- the scalar quanta in this case --- to perform the
clock synchronization.

\begin{figure}[t]
\centering
\includegraphics[width=0.45\textwidth,clip]{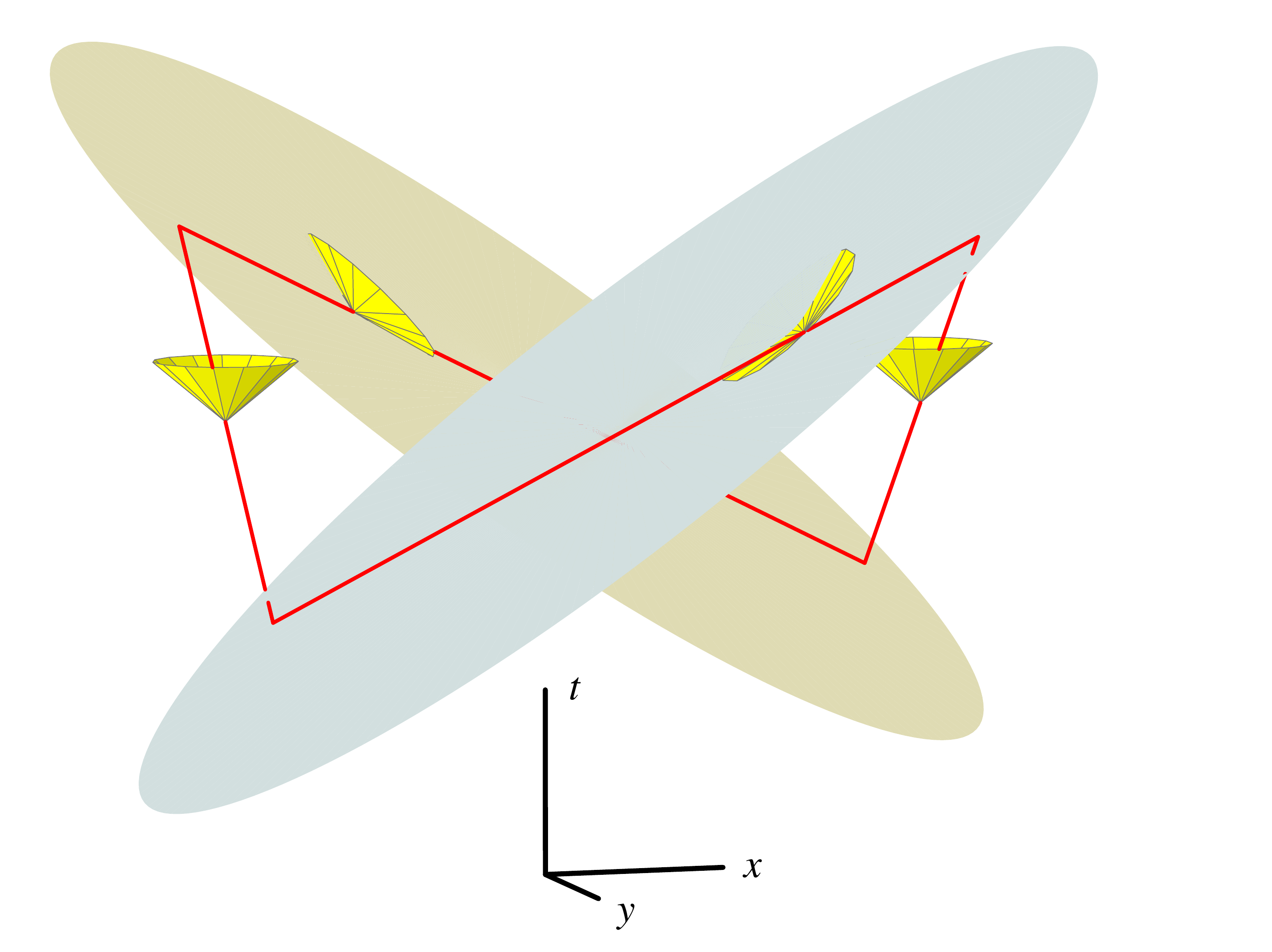}
\caption{\small\it Two regions with superluminal propagation give rise to a closed time-like curve
when boosted towards each other.
}
\label{closed}
\end{figure}

However, if one takes now a pair of such bubbles and boosts them in
the opposite directions, one obtains a field configuration with a
closed time-like curve (see Fig.~\ref{closed}). At large enough
transverse separation between bubbles all local Lorentz invariant
quantities are small, so that this field configuration is within the
regime of validity of the effective field theory. One may object that this is not a
solution of the field equations; however, there
appears to be no natural constraint on the sources that would forbid
causal pathologies\footnote{This is different from the case of
general relativity where the null energy condition is a natural candidate to prevent a
formation of closed time-like curves starting from the
configurations approaching flat space-time in the infinite past
\cite{Visser:1998ua}.}. These arguments indicate that UV completion of
superluminal effective field theories, if any, 
should be very different  from known quantum field theories
or string theory.

One logical possibility to get around both problems, would be to
consider a theory where the Lorentz invariant state with $X=0$ is
absent whatsoever. Problems related to the analytical properties of
the scattering amplitudes disappear simply because the Lorentz
invariant S-matrix does not exist any longer.  Also the above
construction of the closed time-like curve heavily relies on the
coexistence of Lorentz invariant and Lorentz breaking phases. For
instance, if due to some reasons the whole region where $X\leq 0$ is
never accessible, so that $\d_\mu\phi$ is necessarily time-like (both
with respect to the Minkowski and to the acoustic metric), one can always
choose $\phi$ itself as a global time, so that no causal paradoxes
arises. Clearly, none of the known UV complete
quantum theories exhibit this behavior.
 
One of the motivations to consider two-dimensional models, is that in
two dimensions there are better chances to construct renormalizable
theories with this kind of behavior. 
Indeed, in Sec. \ref{sec:aether} we will present a
model (of a vector field, rather than of a scalar) that at the
classical level does not possess a Lorentz-invariant vacuum
state. A well-known feature of two-dimensional physics is
that continuous symmetries that appear to be broken spontaneously at
the classical level are restored in the quantum theory at long
distances. Still, even in the presence of the
Poincare-invariant vacuum one can get around the
above arguments against superluminality due to peculiar properties of
two-dimensional 
space-time.

\subsection{Life is special in two dimensions}  
\label{sec:2d}
First, let us see why the argument about closed time-like curves is
not applicable in two dimensions. 
Let us again consider superluminal
effective theories of the form (\ref{Xaction}).  For simplicity, let
us concentrate on {\it strictly} superluminal theories, {\it i.e.}
such that $P''\leq 0$ for all values of $X$. 
If conditions (\ref{goodness_condition}), ensuring that an
effective field theory is free of local pathologies, are satisfied
(this is always the case in the vicinity of the Lorentz-invariant
vacuum $X=0$, if this vacuum itself is free of ghosts, $P'(0)>0$),
then locally the metric (\ref{effective_metric}) defines a regular
causal structure. The vector $t^\mu=(1,0)$ is time-like at each point
with respect to the effective metric, and points inside the future
light cone. Assume now that there exists a closed curve which is
everywhere time-like and future-directed with respect to the metric
(\ref{effective_metric}). 
Without loss of generality one can assume that this curve has no self-intersections, as all self-intersections can be smoothly deformed out while keeping the curve time-like.
But
for any closed oriented curve without self-intersections on a plane one can find at least one
point where its tangent vector is collinear to the past time-like vector
$-t^\mu=(-1,0)$. Thus we arrive at the contradiction which means that 
it is impossible to construct a closed time-like curve.
In more physical terms closed time-like curves are absent because
for any setup of the type shown in
Fig.~\ref{closed} the two bubbles will necessarily collide as in 2d there is
no possibility to separate them in the transverse direction. As a result
the effective field theory will break down.
 
This argument suggests that superluminal travelers may have better
chances to succeed in two space-time dimensions. What is even more
encouraging is that, as we are going to see, in two dimensions
superluminal travel is possible even in the state with {\it unbroken}
Lorentz invariance, provided the space parity is broken. This will be
the key observation exploited in the current paper.
 
\begin{figure}[t]
\centering
\includegraphics[viewport=00 320 600 520,width=1.0\textwidth,clip]{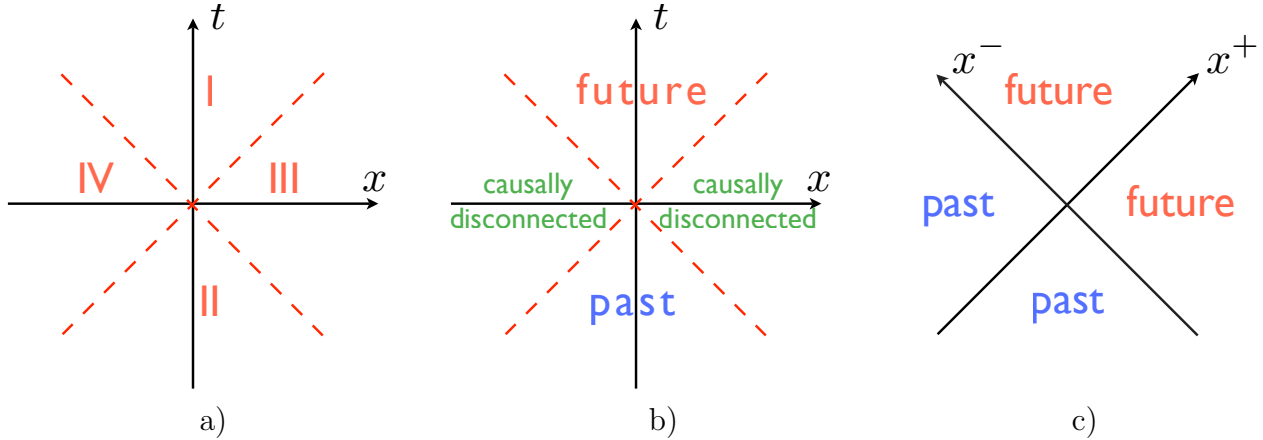}
\put(-410,5){a)}
\put(-240,5){b)}
\put(-80,5){c)}
\caption{\small\it{\rm a)} Light cones emanating from a given point divide the two-dimensional space-time into four 
separate regions. As a result two causal orderings are possible: {\rm b)} the standard  and {\rm c)} the $x^+$-ordered causal structure.
 }
\label{1234}
\end{figure}

Indeed, in two dimensions light cones emanating from a given point
divide the space-time into {\it four}, instead of three in higher
dimensions, separate regions (see Fig.~\ref{1234}): interiors of the
future and past light cones (regions I and II) and two space-like
regions (III and IV). 
 The standard causal structure assumes that the
region I is the causal future of the origin $O$, the region II is the
causal past, while regions III and IV are causally disconnected from
$O$.  Clearly, this choice is compatible with the extended Poincare
group including also the spatial parity transformation $x\to -x$.  The
latter is necessary to map points from the region III into the region
IV.  Consequently, if the spatial parity is broken, there is no reason
for these two regions to be in the same causal relation with
$O$. Interestingly, there is another possible choice of a causal
structure, which is fully compatible with a connected part of the
Poincare group. Namely, one may choose the union of regions I
and III to form the causal future of the origin and the union of
regions II and IV to form the causal past. There is no causally
disconnected region in this case, so that this choice of causal
structure corresponds to the presence of instantaneous interactions.
 
A peculiar property of this causal structure is that it uniquely fixes
the way the space-time is foliated by Cauchy surfaces. These are the
surfaces of a constant light-cone variable
\[
x^+=\frac{1}{\sqrt{2}}(t+x)\;.
\]
Consequently this variable plays the role of a global time, and it is
natural to call the causal structure introduced above the
``$x^+$-ordered'' causal structure. The
second light cone variable
\[
x^-=\frac{1}{\sqrt{2}}(t-x)
\]
is a space coordinate. It is amusing that in two dimensions the
existence of a global time is compatible with Lorentz invariance. 

Thus one concludes that in two space-time dimensions superluminal travel is possible even in Lorentz-invariant theories with a fixed causal structure. The study of this possibility is our main goal in the current paper. 
 
A rather trivial example of a field theory with the $x^+$-ordered
causal structure is 
\be
\label{trivial}
S=-\int d^2x\,(\d_-A^-)^2\;.
\ee
This action is Poincare invariant with $A^-$ transforming as a $(-)$
component of a vector. When quantized in the light-cone coordinates,
using $x^+$ as the time variable, the field $A^-$ mediates an
instantaneous linear potential. Of course this model is rather
trivial as it does not describe any propagating degrees of freedom. To
bring more life into this model we need to couple $A^-$ to a truly
dynamical degree of freedom having a non-trivial $x^+$ evolution.

One way to achieve this is to write an action of the form
\be
\label{lesstrivial}
S=\int d^2x\left[-(\d_-A^-)^2+(\d_+A^+)^2-m^2(A^-A^+)^2\right]\;.  
\ee
To gain some insight into the physics of this system note that this
action does not involve time ($=x^+$) derivatives of $A^-$, so this
field can be integrated out.  The integral over $A^-$ is Gaussian, so
this procedure reduces to calculating a determinant and results in a
highly non-local (in ``space'' coordinate $x^-$) interaction.
It would be interesting to study whether (\ref{lesstrivial}) or its
generalizations provide examples of consistent interacting
theories. One may hope that this is the case, given that the Wick rotation
$x^+\to -ix^+$ transforms (\ref{lesstrivial}) into the positive
definite Euclidean action.  Note that a mass term $\mu^2A^-A^+$ would
spoil this property and introduce instabilities. However, such a term
may be forbidden by a discrete symmetry $A^-\mapsto -A^-$.
 
Note also, that the conventional Wick rotation $t\to -i t$ would not
give rise to a positive definite Euclidean action.  More generally, a
necessary prerequisite for the superluminal propagation is that it
should not be possible to perform a conventional Wick rotation
resulting in a real positive definite Euclidean action.  Indeed,
superluminal propagation means that a retarded propagator of a certain
field $\phi$ does not vanish outside the light cone.  Equivalently, the
$T$-ordered two point function of $\phi$ has an imaginary part outside
the light cone.  If one could make a conventional Wick rotation
resulting in the positive real action, it would be possible to
calculate the $T$-ordered product outside the light cone directly in
the Euclidean space with a manifestly real result
\cite{Dubovsky:2007ac}.
 
In spite of  the spatial parity breaking, there is
no problem to rewrite the action (\ref{lesstrivial}) in the covariant form by
making use of the antisymmetric Levi--Civita tensor
$\epsilon^{\mu\nu}$, and introduce coupling to gravity,
\be
S=\int d^2x\,\sqrt{-g}
\left\{-\nabla_\mu A^{\mu}\frac{\epsilon^{\nu\lambda}}{\sqrt{-g}}\nabla_\nu
  A_\lambda -\frac{m^2}{4}\left(g_{\mu\nu}A^\mu A^\nu\right)^2\right\}
\ee

In the current paper we proceed to a somewhat more involved set of
models exhibiting the same causal structure.  One reason is that,
as we hope to convince the reader, these models are interesting on
their own right.  A more concrete motivation is that superluminal
models of the type (\ref{lesstrivial}) may appear to be disconnected
from theories exhibiting (approximately) conventional behavior. Namely,
it seems impossible to couple them to a dynamical sector
with conventional properties, such that superluminal effects in this
sector would be just a small correction. Indeed, let us consider a
conventional free massive scalar field,
\[
S=\frac{1}{2}\int d^2x \left[(\d\phi)^2-m^2\phi^2\right]\;.
\]
In the light cone coordinates 
the corresponding field equation is first order in ``time'' $x^+$,
\be
\label{simplescalar}
2\d_+\d_-\phi+m^2\phi=0\;.  
\ee 
Consequently, the initial value problem for a massive scalar field
formulated on the set of slices
$x^+=const$ contains only one free function instead of two in the
conventional Cauchy problem formulated in the coordinates
$(t,x)$. This, in turn, leads to inequivalent canonical quantization of
the theory for the two choices of coordinates\footnote{The
  situation is different in the case of a massless field where the
  second free function --- the left-moving part of the field ---
appears as  a gauge degree of freedom due to the presence of  the first class constraints.}.  
On the other hand, as discussed above,
constant $x^+$ surfaces are the only possible choice of the Cauchy
surfaces for the instantaneous systems such as (\ref{lesstrivial}). So
it appears impossible to couple (\ref{lesstrivial}) to a conventional
massive field while keeping instantaneous effects small.

To get around this problem we suggest the following. Let's assume that
there is a dynamical vector $V^\alpha$, such that its $(+)$ component
is non-vanishing.  Then one can modify the scalar field action in
the following way
\be
\label{scalmodified}
S=\frac{1}{2}\int d^2x\left[(\d\phi)^2-m^2\phi^2
+\alpha(V^+\d_+\phi)^2\right]\;,
\ee
where $\alpha$ is a coupling constant.
This changes the effective metric for the propagation of the field
$\phi$, so that the surfaces $x^+=const$ are space-like with respect
to the new 
metric. 
As a result, the scalar field equation becomes second order in
the light cone coordinates, so that the number of degrees of freedom
in the initial value 
problem matches that in the conventional case.
Clearly, for this trick to work we
need $V^+\neq 0$ everywhere for all possible backgrounds.  On the
other hand, in the limit when $\alpha$ is small the system
(\ref{scalmodified}) smoothly approaches the conventional scalar field
(\ref{simplescalar}).  In other words, if both $\alpha$ and a coupling
between the field $\phi$ and the instantaneous sector are small, an
observer coupled just to $\phi$ may pretend for a while that she lives
in a space-time with the conventional causal structure, until effects
mediated by the instantaneous sector force her to change her mind.

These considerations lead us again to the question of finding a theory
with spontaneous Lorentz violation, such that one always has $V^+\neq
0$. Clearly, as formulated, this proposal may appear superficial as it
apparently contradicts to the CMW result on the absence of spontaneous
symmetry breaking in two dimensions.  We will elaborate on this point
later. Remarkably, as we will see now, the theories with $V^+\neq
0$ at different choices of parameters by themselves provide quite non-trivial examples of systems with all possible
causal ordering in two dimensions including the $x^+$-ordered causal structure.


\section{Description of the model and its causal structure}
\label{sec:aether}
At the effective field theory level models developing Lorentz breaking
condensates of a vector field were studied rather extensively (see
e.g.~\cite{Jacobson:2000xp,Gripaios:2004ms,Libanov:2005vu,Gorbunov:2005dd,Cheng:2006us,Eling:2006xg,Gorbunov:2008dj}).
To study a possibility of UV completion of such a model in two
space-time dimensions the most appropriate setup to start with is the
``Einstein-aether'' model of \cite{Jacobson:2000xp,Eling:2006xg}.  This
is a model describing dynamics of a space-time vector field $V_\mu$,
subject to a constraint \be
\label{constraint}
V_\mu V^\mu=1\;.
\ee
In flat space-time the action of this model is of the form
\be
S=\int d^2x\l
-\alpha_1\d_\mu V^\nu \d^\mu V_\nu
-\alpha_2 \d_\mu V^\mu \d_\nu V^\nu
-\alpha_3\d_\mu V^\mu\epsilon^{\nu\lambda}\d_\nu V_\lambda
+\lambda (V^\mu V_\mu-1)\r\;,
\label{EAction}
\ee 
where $\lambda$ is a Lagrange multiplier enforcing the constraint
(\ref{constraint}). In space-time dimensions greater than two one may
also add the term 
$(V^\mu\d_\mu
V^\nu\cdot V^\lambda\d_\lambda V^\nu)$; however, in the two-dimensional
case the latter term can be expressed via the terms already present in
the action (\ref{EAction}) \cite{Eling:2006xg}. The parity breaking
term proportional to $\alpha_3$ was not included in the original
Einstein-aether action \cite{Eling:2006xg}.

The system (\ref{EAction}) is a non-linear sigma model with $SO(1,1)$
symmetry group. This group is one dimensional, so normally the corresponding
sigma-model would just describe a free Goldstone 
field. What makes the situation different in the present case is that the
$SO(1,1)$ group under consideration is a
space-time Lorentz group, rather than an internal symmetry group. This
leads to non-trivial self-interaction of the Goldstone field. To see
this explicitly one solves the constraint (\ref{constraint}) by
introducing the ``rapidity'' field $\psi$ directly parametrizing the
group manifold,
\be
V^{\pm}=\frac{1}{\sqrt{2}}\e^{\pm\psi}\;.
\ee
The spontaneously broken
Lorentz symmetry is realized non-linearly in terms of the Goldstone
field $\psi$,
\be
\label{psitrans}
\psi(x^+,x^-)\mapsto\psi(\e^\gamma x^+,\e^{-\gamma}x^-)+\gamma\;.
\ee
Now the Einstein-aether action (\ref{EAction}) takes the following form,
\be
\label{Spsi}
S=\int dx^+ dx^-
\left\{\frac{1}{g^2}\d_+\psi\d_-\psi
+\frac{\beta_{(+)}}{2g^2}(\d_+\psi)^2\e^{2\psi}+\frac{\beta_{(-)}}
{2g^2}(\d_-\psi)^2\e^{-2\psi}
\right\}\;,
\ee
where the couplings $g^2,\beta_{(\pm)}$ are related to $\alpha_{1,2,3}$ 
in the following way,
\begin{gather}
g^2=\frac{1}{2\alpha_1+\alpha_2}\;,\\
\beta_{(\pm)}=\frac{-\alpha_2\pm\alpha_3}{2\alpha_1+\alpha_2}\;.
\end{gather}
We assume $g^2>0$.
When $\beta_{(+)}=\beta_{(-)}=0$ the action (\ref{Spsi}) becomes free
and acquires an enhanced symmetry: the space-time
Lorentz transformations get disentangled from the shift of the
$\psi$-field. 
In the
vector field language this case corresponds to $\alpha_2=\alpha_3=0$, so
that the second and the third terms in (\ref{EAction}) vanish. 
This enhances the symmetry of the action to $SO(1,1)\times
SO(1,1)$ with the two factors acting independently on the space-time
coordinates and the vector field $V^\mu$. 
Only the second $SO(1,1)$ factor, which plays the role of an internal symmetry, 
is spontaneously broken and the sigma-model gives rise 
to a free theory,  
as it should be.

Our aim is to use the model (\ref{Spsi}) for exploring the
$x^+$-ordered causal structure. The Hamiltonian of the system, conjugate to 
the light-cone variable $x^+$, has the form
\be
\label{Ex+}
H_{lc}=\int dx^-\left\{\frac{\beta_{(+)}}{2g^2}(\d_+\psi)^2\e^{2\psi}
-\frac{\beta_{(-)}}
{2g^2}(\d_-\psi)^2\e^{-2\psi}
\right\}\;.
\ee
A necessary (but not sufficient) condition for the energy to be
positive is $\beta_{(+)}>0$. We adopt this choice in what follows unless stated otherwise.

By power-counting the sigma-model action (\ref{Spsi}) is
renormalizable and, as we discuss below, at least for some values of
the parameters there are good reasons to believe that the theory can
be defined non-perturbatively. Thus, it can be used as a sector giving
rise to 
non-vanishing $
V^+$ in the action (\ref{scalmodified}).

As already mentioned above, existence of non-vanishing $V^+$ 
seems to be at odds with the CMW theorem. To clarify this
issue let us recall that the precise statement of the CMW theorem is
that there cannot exist a {\em symmetry breaking vev} $\langle
V^+\rangle$. This does not imply that the notion of a background
$V^+(x^+,x^-)$ completely looses its meaning. In practice, when
calculating  Green's functions of the system it is often convenient
to separate variables in the path integral into a slowly varying
background part and fast fluctuations. Then, contributions of
the latter into the path integral can be evaluated using the
perturbation theory in the background of the smooth part. The
subsequent integration over the slow modes can be considered as a sort
of ``averaging'' of the perturbative Green's function over all
possible backgrounds. This ``averaging'' leads to restoration of the
symmetry at large distances. Nevertheless, the properties of the 
short distance physics
are captured to large extent by perturbation theory around the
symmetry breaking background, cf. \cite{Witten:1978qu}.
Thus one concludes that 
when describing the system (\ref{scalmodified}) at short time and
distance scales one can use perturbation theory around some particular
value of $V^+$ and, as a consequence, the constant $x^+$ surfaces
provide a well-defined set of Cauchy slices. Roughly, at long
distances one should ``average'' over all possible configurations of 
$V^+$ , but as
$x^+$ is a good time variable for every individual background $V^+$,
this should remain the case after making the ``average''.

\begin{figure}[t]
\centering
\includegraphics[width=0.3\textwidth,clip]{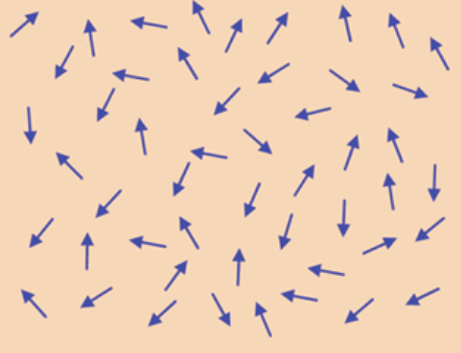}
\caption{\small\it The net magnetization is averaged out to zero at long distances in a two-dimensional ferromagnetic.
Nevertheless, a perturbative description in terms of small excitations around symmetry breaking configurations is appropriate for many local questions. 
}
\label{spins}
\end{figure}

It is worth discussing the physical meaning of the CMW result also
from the point of view of the canonical quantization. 
In the context of condensed matter
systems one may think, for instance, about ferromagnetic material.
Then, performing a detailed measurement, one finds 
locally at each point of space a spin pointing in
some direction (see Fig.~\ref{spins}). The CMW statement
is that in two dimensions, when one describes the physics of such a
system at long distance scales, fluctuations of spins lead to
``averaging'' of the local spins over all possible directions and result
in restoration of the symmetry. Similarly, in the
context of the Minkowski space field theory, long range quantum
fluctuations result in the symmetry restoration when one considers
asymptotic quantities such as the $S$-matrix elements. On the other
hand, the quantum theory retains many properties of the classical
description which operates with the notion of a given field
configuration. In particular, in what follows it will be important for
us that the quantum theory inherits the causal structure of the
classical one (see \cite{Dubovsky:2007ac} for a recent discussion).

With these comments in mind let us proceed to the inspection 
of the model (\ref{Spsi}). One notices 
that the constant shift of the Goldstone field
$\psi\to\psi+\gamma$ changes the ratio $\beta_{(-)}/\beta_{(+)}$ by an
arbitrary positive multiplicative factor, so it is only the product
$\beta_{(+)}\beta_{(-)}$ which has a physical meaning. There are
four different cases to consider:
\subsection{$\beta_{(+)}=-\beta_{(-)}\equiv\beta$}

\begin{figure}[t]
\centering
\includegraphics[viewport=00 120 600 690,width=0.353\textwidth,clip]{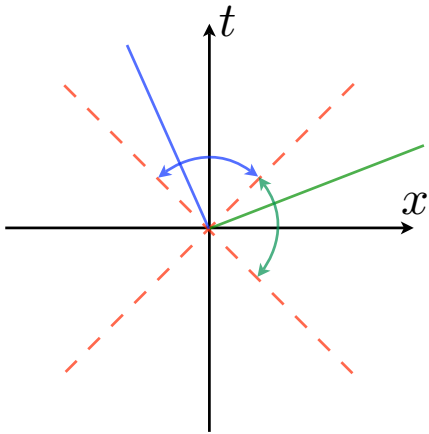}
\caption{\small\it For $\beta_{(+)}=-\beta_{(-)}$,  depending on the value of $\psi$,  the left-mover can propagate in any direction inside the future light cone and the right-mover can propagate  
 in an arbitrary
direction in the right space-like region.}
\label{averaging}
\end{figure}

This choice of parameters explicitly breaks the spatial parity
$x\mapsto -x$. The light-cone Hamiltonian (\ref{Ex+}) is positive-definite
in this case and is minimized at the classical level by configurations
$\psi=const$. We will see shortly that this theory naturally obeys the
$x^+$-ordered causal structure. However, to make a comparison with the
conventional approach more transparent let us see what happens if one
tries to use $t$ and $x$
as the time and space variables. Let us fix the perturbative vacuum
$\psi=0$. Around this vacuum one
finds two modes: left and right movers, with
dispersion relations,
\be
\label{dis}
\omega_{L,R}=v_{L,R}k_{L,R}\;.
\ee
As a result of parity breaking the two propagation velocities are not
equal,
\be
v_R=\beta+\sqrt{1+\beta^2}\;,\;\;v_L=-\beta+\sqrt{1+\beta^2}\;.
\ee
Importantly, the right-moving particle propagates superluminally,
while the left-moving one subluminally. Taking a different value of
$\psi$ is equivalent to looking at this situation from the point of
view of a boosted observer. By choosing an appropriate value of
$\psi$ one may obtain a left-mover propagating in an arbitrary
direction in the future light cone. Similarly, with an appropriate value
of $\psi$ on can obtain a right-mover propagating in an arbitrary
direction in the right space-like region (see
Fig.~\ref{averaging}). 
Of course, for large enough value of $\psi$ the
right-mover propagates back in the Minkowski time $t$, indicating that
the well-defined time variable is actually $x^+$ rather than $t$.
Using again the intuition that the true long distance correlators can
be thought of as a result of ``averaging'' over different perturbative
vacua, we see that the retarded Green's function of the theory is
non-zero in the whole region\footnote{Note that as a result of the
  ``averaging'' one expects the complete Green's functions of the
  system to be Lorentz-invariant, in spite of the fact that the
  perturbative modes have Lorentz-violating dispersion relations
  (\ref{dis}). This is somewhat analogous to the situation in QCD
  where the asymptotic states are colorless in spite of the fact that
  the short distance dynamics is described by colored quarks.} 
$x^+>0$. Thus, this theory enjoys the
$x^+$-ordered causal structure.  


\subsection{$\beta_{(-)}=0$}
\label{0case}
This is also a parity breaking case. The light-cone Hamiltonian
(\ref{Ex+}) is again positive, though now the ground state is highly
degenerate at the classical level: any configuration $\psi=\psi(x^-)$
has zero energy. This degeneracy is related to a large symmetry of the
action (\ref{Spsi}) with $\beta_{(-)}=0$, discussed in
Sec.~\ref{beta0}.

\begin{figure}[t]
\centering
\includegraphics[width=0.78\textwidth,clip]{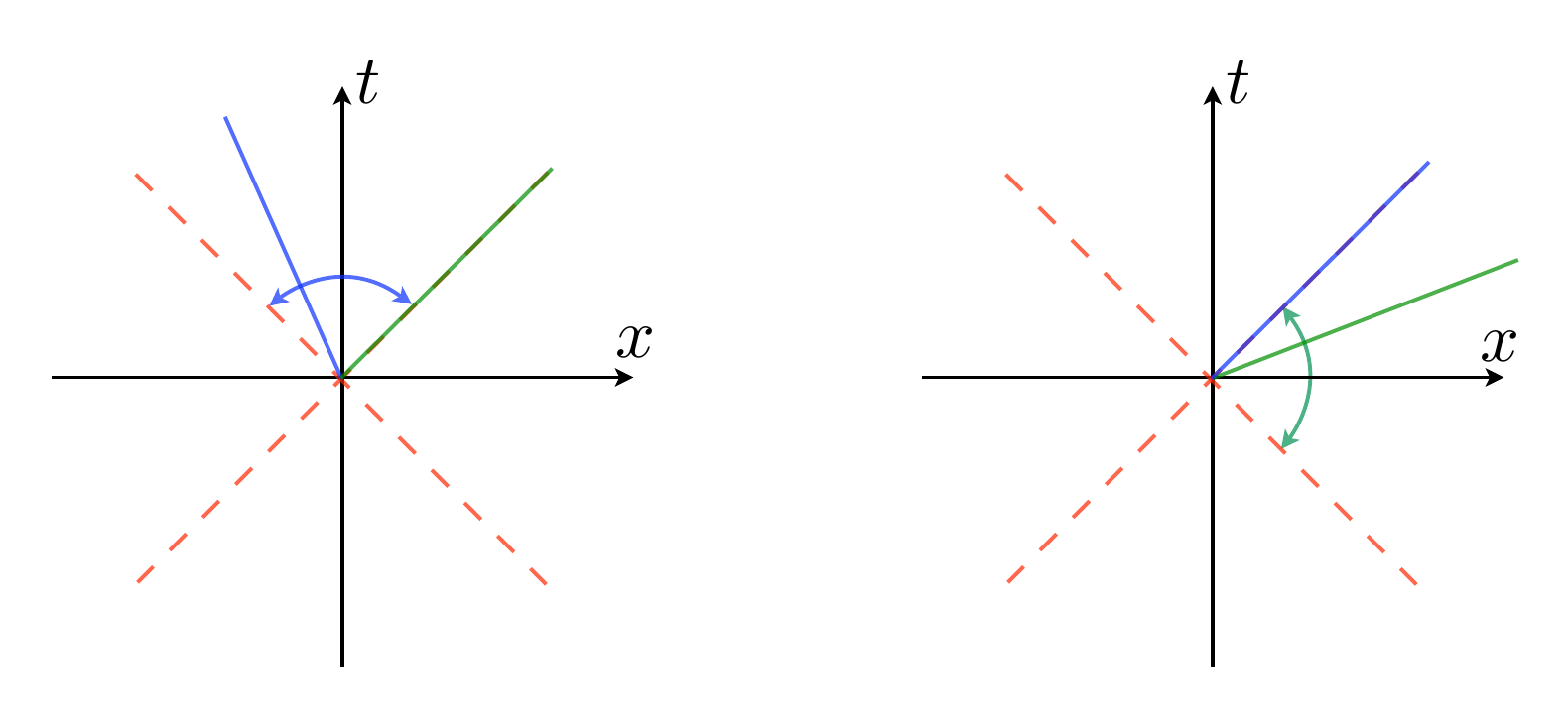}
\put(-330,5){a)}
\put(-110,5){b)}
\caption{\small\it For $\beta_{(-)}=0$, one of the modes propagates along the light cone.
Depending on the sign 
of the $\d_+\psi\d_-\psi$ term 
 the other mode propagates
either inside the future light cone {\rm a)} or in right space-like region {\rm b)}.}
\label{fig:sigma}
\end{figure}

In spite of the parity breaking, the theory (\ref{Spsi}) with
$\beta_{(-)}=0$ by itself possesses
the standard causal structure. 
To see this let us again work in terms of the coordinates $t$, $x$. 
Around the
vacuum $\psi=0$ one finds 
two modes: a right-mover propagating with
the speed of light, and a subluminal mode propagating with the
velocity $({\beta-2})/({\beta+2})$, where $\beta\equiv\beta_{(+)}$. This mode
is left-moving for $\beta<2$
and right-moving for $\beta>2$.  A variation  of $\psi$ is equivalent to a variation of $\beta$ and results
in changing the velocity of the subluminal mode
that can propagate in any direction inside of the light cone,
depending on the value of $\psi$ (see Fig.~\ref{fig:sigma}a). Consequently, both at the classical and quantum  levels
this theory possesses the standard causal structure.

Still, by a slight modification of the setup
it is easy to construct theories obeying the $x^+$-ordering.
One introduces a scalar field $\chi$ and couples it to $\psi$ as
follows,
\be
\label{Spsichi}
S=\int dx^+ dx^-
\left\{\frac{1}{g^2}(\d_+\psi\d_-\psi-\d_+\chi\d_-\chi)
+\frac{\beta}{2g^2}(\d_+\psi)^2\e^{2\psi}+
\frac{\beta_\chi}{2g^2}(\d_+\chi)^2\e^{2\psi} \right\}\;, 
\ee 
where $\beta_\chi>0$. The resulting theory is well-defined, in
particular, it has positive light-cone Hamiltonian.
In the background $\psi=0$ the
$\chi$-excitations contain a right-moving mode propagating
with the speed of light and another propagating superluminally (see
Fig.~\ref{fig:sigma}b). The interaction between the fields $\chi$ and
$\psi$ couples the subluminal ($\psi$) and superluminal ($\chi$) modes
leading to a theory with the $x^+$-ordered causal structure.

The theory (\ref{Spsichi}) has very interesting properties that suggest that this model (possibly with some modifications) may be
exactly solvable. 
This issue is discussed in
more detail in Sec.~\ref{beta0}.

\subsection{$\beta_{(+)}=\beta_{(-)}\equiv\beta>0$}
\label{sublcase}
This choice preserves the spatial parity $x\mapsto -x$.  
The
dispersion relation for linear fluctuations around $\psi=0$ 
background has the form
\be
\label{dispersion}
\omega^2=v^2k^2
\ee
with the effective  propagation velocity $v$ given by
\be
\label{cpsi}
v^2=\frac{1-\beta}{1+\beta}\;.
\ee
For $\beta>0$ the perturbations are subluminal. According to the
arguments above this implies that the model has the standard causal
structure. 
The system is perturbatively stable 
for $|\beta|<1$. However, the light-cone Hamiltonian (\ref{Ex+}) is
unbounded from below. The same is true for the Hamiltonian conjugate
to the normal time $t$, see Eq.~(\ref{hamiltonian}) below. As will
be discussed in more detail in Sec.~\ref{sec:equalbeta} this  may signal a presence of a
non-perturbative instability in the model.
\subsection{$\beta_{(+)}=\beta_{(-)}\equiv\beta<0$}

Note that Eq.~(\ref{cpsi}) implies that the system is perturbatively
stable also for $0>\beta>-1$. In this case the velocity (\ref{cpsi})
is greater than 1, i.e. the perturbative modes are
superluminal. This last case is rather different from 
all the previous ones, and
is somewhat outside the main line of the current paper.
Indeed, existence of the
superluminal modes propagating in both directions cannot be reconciled
with the restoration of Lorentz symmetry required by the CMW
theorem if the theory has any fixed causal structure. 
One possible attitude at this point could be that this model is
inconsistent.

However, in principle, quantum field theories without fixed causal
structure might exist (gravity being one example). 
Below, we
try to take
an optimistic viewpoint that the theory (\ref{Spsi}) makes sense also in this case
and see where this assumption may lead us. The bulk of the discussion
is postponed until Sec. 5.4. Here we limit ourselves to the
observation that at the classical level the theory admits a well-posed 
initial value problem provided
one uses surfaces that are space-like with respect to the effective
metric determining propagation of small perturbations (see eq.~(\ref{Gpm}) below)  to define the set of Cauchy slices. Note that the
argument of Sec.~\ref{sec:2d} about the absence of closed time-like
curves
applies to the model at hand as well,
so that the initial value problem is well-defined not only locally but
also globally.

\section{One-loop running}
\label{sec:oneloop}

To gain more insight into the dynamics of the Einstein-aether theory 
it is instructive to study
the one-loop running of the parameters $g^2,\beta_{(\pm)}$. This can be
done in a uniform way for all the choices of $\beta_{(+)}$,
$\beta_{(-)}$. 

To get around the infrared divergencies it is convenient to make use
of the background field method (cf., e.g., the
classic calculation \cite{Polyakov:1975rr}).
In this approach one decomposes the field $\psi$ into a long-distance
part $\psi_0$ with momenta smaller than some value $\tilde\Lambda$ and
the fluctuating part $\pi$ with momenta greater than
$\tilde\Lambda$. The fluctuations $\pi$ are then integrated out for
the range of momenta between $\tilde\Lambda$ and the UV cutoff
$\Lambda>\tilde\Lambda$. 
The Lorentz symmetry (\ref{psitrans}) implies that as a result of this
procedure one will obtain the action for $\psi_0$ 
of the form (\ref{Spsi}) with
couplings evaluated at the scale $\tilde{\Lambda}$ plus higher
dimensional irrelevant interactions.

\begin{figure}[t]
\centering
\includegraphics[width=0.4\textwidth,clip]{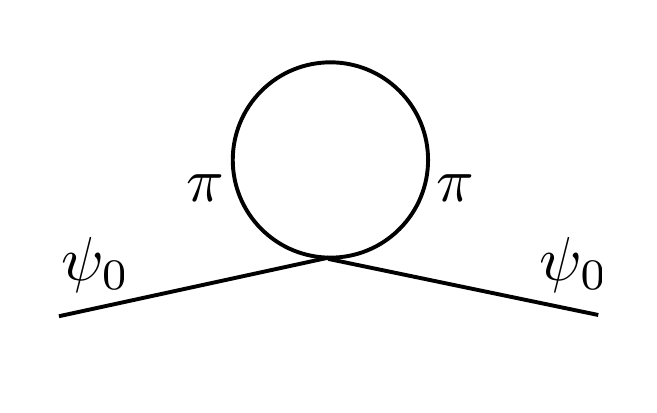}
\caption{\small\it The diagram contributing to one-loop running.}
\label{diagram}
\end{figure}

At the one loop level we need only the part of the
action  quadratic in $\pi$ (given that we are not interested in keeping track of the
irrelevant operators). The latter is conveniently presented in the
following form,
\be
\label{S2pi}
S^{(2)}[\psi_0,\pi]=\frac{\sqrt{1-\beta_{(+)}\beta_{(-)}}}{2g^2}\int d^2x 
\sqrt{|G|}\l G^{\mu\nu}\d_\mu\pi\d_\nu\pi-{\cal
M}^2\pi^2\r\;,
\ee
where the metric $G^{\mu\nu}$ is given by
\be
\label{Gpm}
G^{++}=\beta_{(+)}\e^{2\psi_0}\;,\;G^{--}=\beta_{(-)}\e^{-2\psi_0}\;,\;G^{+-}=1
\ee
and the mass ${\cal M}^2$  is
\be
\label{M2}
{\cal M}^2=2\beta_{(+)}\e^{2\psi_0}\l(\d_+\psi_0)^2+\d_+^2\psi_0\r
+2\beta_{(-)}\e^{-2\psi_0}\l (\d_-\psi_0)^2-\d_-^2\psi_0\r\;.
\ee
The one loop contribution to the
action $\delta S_{1-loop}$ is given by
\be
\label{1loop}
\e^{i\delta S_{1-loop}[\psi_0]}=
\int{\cal D}\pi\,\e^{iS^{(2)}[\psi_0,\pi]}\;,
\ee
where the integral is taken over modes with momenta
between $\tilde{\Lambda}$ and $\Lambda$.
Note that $\delta S_{1-loop}$  
is a diff invariant functional of the metric $G^{\mu\nu}$ and of the 
mass ${\cal M}^2$. Therefore,
it has the form
\be
\label{shouldbe}
\delta S_{1-loop}[\psi_0]=\int d^2x\sqrt{|G|}(c_0-c_1{\cal{M}}^2+\dots)\;,
\ee
where dots stand for higher dimensional irrelevant operators and
$c_0$, $c_1$ are some coefficients. 
We do not include the Einstein-Hilbert term
$\sqrt{|G|}R_G$  into (\ref{shouldbe})  as this term is a total derivative in two dimensions. Next,
$\sqrt{|G|}$ is just
a constant that doesn't depend on $\psi_0$. Consequently, the only non-trivial running is related to the second term in (\ref{shouldbe}).
By comparing the expression
(\ref{shouldbe}) to the original action (\ref{Spsi}) we find that
$g^2$ remains constant at the one-loop level. To reconstruct the
one-loop RG 
equations for $\beta_{(\pm)}$ we need to calculate the coefficient
$c_1$. It is given by the diagram of Fig.~\ref{diagram},
\be
\label{c1mink}
c_1=\frac{\sqrt{1-\beta_{(+)}\beta_{(-)}}}{2g^2}
\int\frac{d^2p}{(2\pi)^2}\langle\pi(p)\pi(-p)\rangle\;,
\ee
where
the $\pi\pi$ propagator is determined by the local value of the $G^{\mu\nu}$ metric,
\be
\label{pipi}
\langle\pi(p)\pi(-p)\rangle=\frac{ig^2}{2p_+p_-+\beta_{(+)}\e^{2\psi_0}p_+^2
+\beta_{(-)}\e^{-2\psi_0}p_-^2+i\epsilon}\;.
\ee
Note that the $i\epsilon$-prescription in (\ref{pipi}) is chosen to be
compatible with the $x^+$-ordered causal structure. It agrees with
the conventional  time-ordering in the cases considered in Secs.~\ref{0case}, \ref{sublcase}.

To evaluate the integral (\ref{c1mink}) it is convenient to make an
analytic continuation 
\[
p_{\pm}\mapsto\pm\frac{p}{\sqrt{2}}\e^{\pm i\phi}\;.
\]
This continuation is nothing else as the standard Wick rotation
$p_0\mapsto -ip_0$. Note that due to non-standard structure of
the propagator (\ref{pipi}) the integration contour generically hits
poles in the complex $(p_+,p_-)$ space during the Wick rotation, and
one should be careful to properly include contributions of these poles.
This complication does not arise for sufficiently small values of
the Lorentz-violating parameters $\beta_{(\pm)}$. Thus, the simplest
way to perform the integral is to evaluate it for small $\beta_{(\pm)}$
and then analytically continue to larger values of $\beta_{(\pm)}$. 
One obtains
\be
\label{c1loop}
c_1=\frac{\sqrt{1-\beta_{(+)}\beta_{(-)}}}{8\pi^2}
\int_{\tilde{\Lambda}}^\Lambda\frac{dp}{p}\int_0^{2\pi}
d\phi\left[1-\frac{\beta_{(+)}\e^{2\psi_0}}{2}\e^{2i\phi}
-\frac{\beta_{(-)}\e^{-2\psi_0}}{2}\e^{-2i\phi}\right]^{-1}=
\frac{1}{4\pi}\log{\frac{\Lambda}{\tilde{\Lambda}}}\;.
\ee
Comparing this result to the original 
action for $\psi$ (\ref{Spsi}) we find 
\[
\beta_{(\pm)}(\tilde{\Lambda})=\beta_{(\pm)}(\Lambda)
+\frac{g^2\beta_{(\pm)}(\Lambda)}{\pi\sqrt{1-\beta_{(+)}\beta_{(-)}(\Lambda)}}
\log{\frac{\Lambda}{\tilde{\Lambda}}}\;.
\]
Equivalently, couplings $\beta_{(\pm)}$ satisfy the following RG equations
\be
\label{RGE}
\frac{d\beta_{(\pm)}}{d\log{\Lambda}}=
-\frac{g^2\beta_{(\pm)}}{\pi\sqrt{1-\beta_{(+)}\beta_{(-)}}}\;,
\ee
These equations (\ref{RGE}) imply that the ratio
$\beta_{(-)}/\beta_{(+)}$ remains constant while the product 
$\beta_{(+)}\beta_{(-)}$
flows to zero in the UV.
Consequently, at high energies both
Lorentz violating parameters $\beta_{(\pm)}$ flow to zero and the theory
is asymptotically free\footnote{Strictly speaking, this statement  is correct only
  for the cases  when both
  $\beta_{(\pm)}$ are non-zero. 
In the $\beta_{(-)}=0$ case $\beta_{(+)}$ also flows to
zero in the UV, however running of $\beta_{(+)}$ does not have
physical meaning in this case, as the value of
$\beta_{(+)}$ can be changed by a shift of $\psi$.}.  

At this point it is worth mentioning that the Euclidean $SO(2)$
version of the model (\ref{Spsi})  with $\beta_{(+)}=\beta_{(-)}$, whose action is
obtained by substituting $\psi\to i\psi$ and $g^2\to-g^2$, describes
nematic liquid crystals \cite{nematics}. The corresponding RG equation
agrees with (\ref{RGE}) after reversing the sign of $g^2$. However,
this changes the sign of the beta-function, so that in the liquid crystal case
$\beta_{(\pm)}$ are irrelevant parameters. Thus 
the physics of two
models is very different.

Note that the RG flow (\ref{RGE}) is derived in the leading order in
$g^2$ but to all orders in $\beta_{(\pm)}$. There is no reason to expect
that the coupling $g^2$ remains constant at the higher loop 
level\footnote{Except the case $\beta_{(-)}=0$ where $g^2$ is not
  renormalized in 
 all orders of the perturbation
theory, see below.}.
Still, the fact that the model with $\beta_{(\pm)}$ 
is free at arbitrary values of $g^2$ implies 
that the $g^2$ running is proportional to
$\beta$. This means that the theory will remain asymptotically free
once the RG flow enters into the region of small $g^2$ and $\beta$. A
straightforward analysis of possible two-loop corrections shows that
this always happens if the RG flow starts at a given scale from any
value of $\beta$ and sufficiently small $g^2$.
Nevertheless, it is possible that the structure of the RG
flow changes qualitatively at larger (but, perhaps, still
perturbative) values of $g^2$. It would be of interest to perform the
two-loop calculation in this model to see if this is
the case.

\section{Nonperturbative considerations}
\label{sec:brokenparity}
Let us now turn to the non-perturbative aspects of the Einstein-aether
model. These are quite different in the four cases listed in
Sec.~\ref{sec:aether}.

\subsection{Case $\beta_{(+)}=-\beta_{(-)}\equiv\beta$}
\label{sec:negativebeta}
As already noticed above, this case obeys the $x^+$-ordered
causal structure. Thus, it is natural to interpret $x^+$ and $x^-$ as
time and space variables. The
Wick rotation
$x^+\to -ix^+$ results in the Euclidean action with a positive real
part, strongly suggesting that the theory is consistent at the 
non-perturbative level. Also, the light-cone Hamiltonian (\ref{Ex+})
describing
evolution in $x^+$ is positive definite in this case, 
so the system is stable, 
at least at the semiclassical level.

\begin{figure}[t]
\centering
\includegraphics[viewport=00 220 600 590,width=0.4\textwidth]{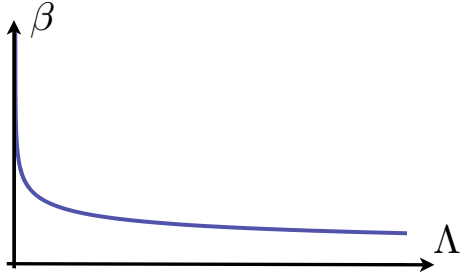}
\caption{\small\it One-loop running for $\beta_{(+)}=-\beta_{(-)}$.}
\label{fig:BKTRGE}
\end{figure}

The RG equation (\ref{RGE}) implies that $\beta$ evolves from zero in
the UV to infinity in the IR (see Fig.~\ref{fig:BKTRGE}). 
This does not pose
any problem as the
perturbation theory
works at all values of $\beta$.
In principle, the IR behavior could change at the higher loop level,
where we expect $g^2$ coupling to run as well. Then $g^2$ could blow
up at some IR scale as a result of this running. 
However, at least in some range of parameters, this
doesn't happen. Indeed, in the limit when $\beta$ is large one can
neglect the first term in the action and arrive at the action of the
form, 
\be
\label{Spsikappa}
S=\frac{1}{2\varkappa^2}\int dx_+ dx_-
\left[(\d_+\psi)^2\e^{2\psi}-(\d_-\psi)^2\e^{-2\psi}\right]
\;,
\ee
where $\varkappa^2=g^2/\beta$. This theory is weakly coupled at small
$\varkappa^2$ and the one-loop RG equation for $\varkappa^2$ can be
obtained from (\ref{RGE}) by taking the limit $\beta\to \infty$,
\be
\label{kappaRGE}
\frac{d\varkappa}{d\log\Lambda}=\frac{\varkappa^3}{2\pi}\;.
\ee
This equation shows that the theory is free in the IR, so that no 
dynamical IR scale is generated.

In the UV the theory flows to the asymptotically free Lorentz symmetric
point $\beta=0$.
Consequently, this model by itself (even without introducing
couplings to additional fields as in (\ref{scalmodified})) provides an
example of a theory possessing the $x^+$-ordered causal structure in
such a way, that in a certain regime an observer may be tricked into
thinking that the causal structure is the standard one. Indeed, the
smallness of
$\beta$ in the UV implies that results of short distance
experiments can be with a good accuracy described in terms of the
conventional causal structure.  To figure out that the actual causal
structure is different, an observer should either wait for a long
enough time (and/or perform an experiment in a large enough volume),
or make precision measurements of the $\beta$ suppressed effects
at short distances.
Note, that these effects, when interpreted within the standard causal
structure, appear completely non-local (and even acausal).  Also the
true causal structure features instantaneous interactions, that
definitely qualify as a non-local effect.  So this theory provides an
example of a field theoretical model giving rise to the genuinely
non-local physics, in spite of a benign looking and local action. A
somewhat similar behavior is exhibited by four dimensional effective
field theories discussed in \cite{Adams:2006sv}. Interestingly, here
one observes this in, as far as we can tell, a fully consistent
microscopic quantum theory.

\subsection{Case $\beta_{(-)}=0$}
\label{beta0}
We are going to see that this case has a number of 
interesting properties that potentially allow to move on rather
far in the non-perturbative analysis of the model and even give rise
to a hope to obtain a solvable model with $x^+$-ordered causal structure.  The action has the
following form 
\be
\label{Spsi0}
S=\int dx^+ dx^-
\left\{\frac{1}{g^2}\d_+\psi\d_-\psi
+\frac{\beta}{2g^2}(\d_+\psi)^2\e^{2\psi}
\right\}\;.
\ee
It has obvious similarities with that of the
Liouville theory.  Like in the Liouville case, $\beta$ is not a
physically meaningful parameter any longer: by making the field
shift
\[
\psi\to\psi+const
\] 
one can change the value of $\beta$ by an arbitrary positive factor.

An important property of the Liouville theory is its conformal
invariance. The action 
(\ref{Spsi0}) is not invariant under the
full conformal group due to the presence of $(\d_+\psi)^2$ in the
seconf term. Still, at the
classical level, (\ref{Spsi0}) is invariant under right mover's
subgroup of the conformal group and under the global part of the left
mover's subgroup, 
\be
\label{conf}
\begin{split}
&x^+\mapsto f(x^+)~,~~~x^-\mapsto g(x^-)\;,\\
&\psi(x^+,x^-)\mapsto\psi(f(x^+),g(x^-))
-\frac{1}{2}\log{f'(x^+)}+\frac{1}{2}\log{ g'(x^-)}\;,
\end{split}
\ee
where 
\[
f=\frac{ax^++b}{cx^++d}\;,~~~ad-bc=1\;,
\]
and $g$ is a generic function.
As in the Liouville case, the field $\psi$ transforms inhomogeneously
under the conformal transformations (\ref{conf}). We are not able to 
say at the
moment whether the whole symmetry (\ref{conf}) can be preserved at the quantum
level. 

\begin{figure}[t]
\centering
\includegraphics[width=0.9\textwidth,clip]{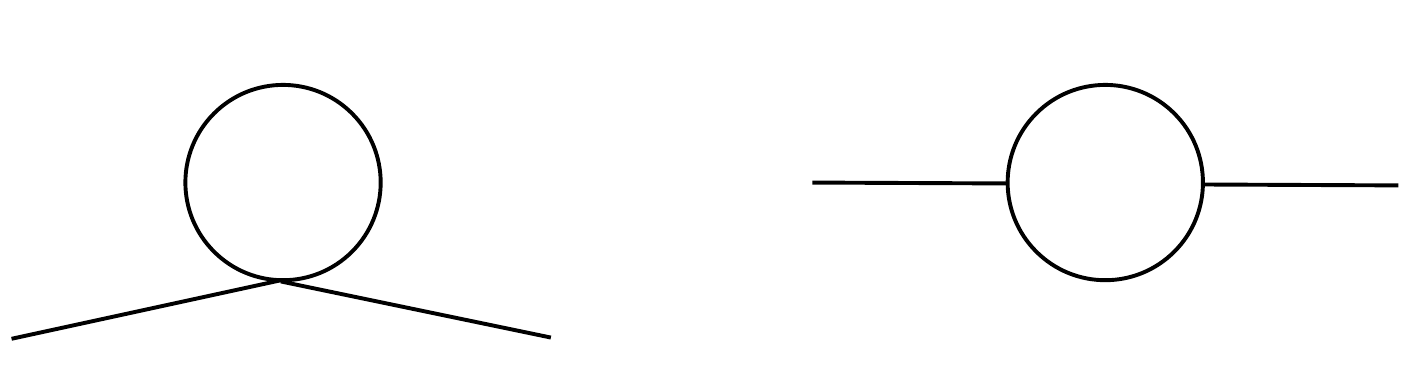}
\caption{\small\it UV divergencies originated from the diagram on the left are removed by the normal ordering.
The right diagram is finite in two-dimensional theories with non-derivative interactions, as well as in a more general class of theories (\ref{Sdgeneral}).}
\label{fig:loops}
\end{figure}

There is yet another property that the model (\ref{Spsi0}) shares with the
Liouville theory. Namely, a simple power
counting shows that in the Liouville theory (and, more generally, in
any two dimensional field theory with non-derivative interactions) the
only UV divergences in Feynmann diagrams are due to loops where an
internal line has both ends on the same vertex
(Fig.~\ref{fig:loops}). As a result, all UV divergencies in these
theories can be removed by the free field normal ordering. In the
Liouville case this allows to perform the canonical quantization of the
theory and to establish a large number of properties (such as the
conformal invariance) at the non-perturbative level
\cite{Curtright:1982gt}.

The action (\ref{Spsi0}) involves derivative interactions, so the 
naive power counting would suggest that there should be divergences
that cannot be removed by the free field normal ordering; for
instance, one would expect the right diagram in Fig.~\ref{fig:loops}
to diverge.
 As we are going to show this conclusion is wrong, and there
is a simple argument involving a modified power counting
that shows that the action (\ref{Spsi0}) can also be
renormalized just by the free field normal ordering.  The
actual statement is even stronger. Namely, we will show that any two
dimensional action of the form 
\be
\label{Sdgeneral}
S=\int dx^+ dx^-
\left[\d_+\psi\d_-\psi
+U(\d_+,\psi)
\right]\;
\ee
is renormalizable by the free field normal ordering. Here $U$ is an
arbitrary function of the field $\psi$ and its derivatives in the
$x^+$ direction.  This result is somewhat unexpected, given that the
naive power counting suggests that, generically, theories
(\ref{Sdgeneral}) are non-renormalizable.  To prove that they are
actually renormalizable, and that all divergences can be removed by
normal ordering, it is convenient to use the language of the Wilsonian
RG. 
Let us assume that with a certain cutoff scale
$\Lambda$ the action takes form (\ref{Sdgeneral}) and let us write the
(generalized) potential function $U$ in the following form 
\be
\label{U}
U=\sum \frac{\alpha_j}{M^{j-2}} U_j(\d_+,\psi)\;,
\ee
where $U_j$ stand for all terms in $U$ that have exactly $j$
derivatives, $M$ is the parameter of dimension mass that gives the
correct dimensionality to all the terms, $\alpha_j$ are dimensionless
parameters, and the functions $U_j$ don't contain any dimensionful
parameters in them.  Note that the action (\ref{Sdgeneral}) is
invariant under Lorentz transformations 
\be
\label{psispur}
\psi(x^+,x^-)\mapsto\psi(\e^\gamma x^+,\e^{-\gamma}x^-)
\ee
if $\alpha_j$ are considered as spurion fields transforming as
\be
\label{alphalaw}
\alpha_j\mapsto\e^{-j\gamma}\alpha_j\;.
\ee
Let us see what restrictions this symmetry implies on the form of the
action obtained by integrating out all modes between $\Lambda$ and
some lower scale $\tilde\Lambda$. Note, that the special
form of the action (\ref{Sdgeneral}) is not protected by any symmetry,
so that in general the effective action at the scale $\tilde\Lambda$ 
will contain terms with 
$\d_-$ derivatives as well. However, the symmetry (\ref{psispur}), (\ref{alphalaw}) and
dimensional considerations
 impose strong restrictions on the dependence of
various terms on the
UV cutoff $\Lambda$. Indeed, let us consider a term 
proportional to a product of $k\geq 1$ coupling
constants $\alpha_{j_1}\cdot ... \cdot \alpha_{j_k}$ in
the effective Lagrangian. The symmetry
(\ref{psispur}), (\ref{alphalaw}) fixes the difference between the
number of $\d_+$ and $\d_-$ derivatives in this term. Moreover, the
dependence of the action 
(\ref{Sdgeneral}) on $M$ implies that this term contains $M^{J-2k}$,
where
$J=j_1+\ldots +j_k$, in the denominator. Requiring that this term  
has mass
dimension 2 one obtains its possible form,
\be
\label{term}
\frac{\alpha_{j_1}\cdot ... \cdot \alpha_{j_k}}{ M^{J-2k}}
\frac{\d_+^J}{\Lambda^{2k-2}}\l\frac{\d_+\d_-}{\Lambda^2}\r^n
\left(\log\frac{\Lambda}{\tilde\Lambda}\right)^m\;.
\ee
We see that the only possible UV
divergences are associated with the terms containing no $\d_-$ ($n=0$),
and proportional to a single coupling constant $\alpha$ ($k=1$). These
are the logarithmic divergencies coming from the loops where an internal line has both
ends on the same vertex as in the left diagram in
Fig.~\ref{fig:loops}, {\it i.e.} exactly those removed by normal
ordering.
 
\begin{figure}[t]
\centering
\includegraphics[viewport=00 170 600 620,width=0.3\textwidth,clip]{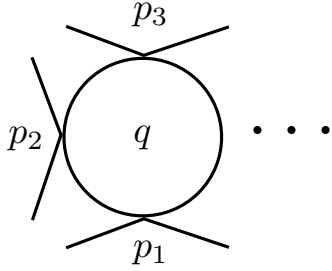}
\caption{\small\it A general one-loop graph.}
\label{fig:genoneloop}
\end{figure}

Given a somewhat formal character of the above argument it is instructive
to see how it works at the diagrammatic level. Let us consider a
general one loop diagram as shown in Fig.~\ref{fig:genoneloop}. 
 After the standard Wick rotation this
diagram gives rise to the Feynmann integral of
the form 
\be
\label{genintegral}
I=\int \frac{d^2q\,q_+^n}{\displaystyle\prod_{i} \l(q+p_i)^2+\mu^2\r}\;,
\ee
where $q$ is the internal loop momentum, $p_i$ is the external
momentum coming from the $i$-th vertex, and we introduced a mass
$\mu^2$ as an IR regulator.  The only difference between this integral
and the usual one obtained 
in the absence of derivative interactions, is the
$q_+^n$ factor in the numerator coming from $\d_+$'s in the
interaction vertices. Naively, this factor changes the converging properties of
the integral by introducing an additional $n$-th power of the momentum
in the numerator.  To see why this naive counting is wrong, it
is convenient to switch to the polar coordinates.
Then the Feynmann
integral (\ref{genintegral}) takes the following form, 
\be
\label{polargenintegral}
I=\int_0^\Lambda dqq^n\int_0^{2\pi}
\frac{d\phi\,\e^{in\phi}}{\displaystyle\prod_{i} \l q^2+p_i^2+\mu^2+2qp
\cos{(\phi-\phi_i)}\r}\;,
\ee
where
\[
q_\pm=q\e^{\pm i\phi}\;,\;\;p_{i\pm}=p_i\e^{\pm i\phi_i}\;.
\]
The point is that $\d_+^n$ brings in not only a factor $q^n$ in the
numerator, but also a phase factor $\e^{in\phi}$. To
see the impact of this factor one Taylor expands the denominator in
(\ref{polargenintegral}) in powers of $\cos{(\phi-\phi_i)}$. At large
$q$ each power of $\cos{(\phi-\phi_i)}$ brings in one factor of
$q^{-1}$. To compensate the phase $\e^{i n\phi}$ and make the angular
integral non-zero one
needs at least $n$ of such factors. 
As a result, after $\phi$ integration one obtains
the integral over $q$ with the same UV behavior as in a theory without
derivative interactions. Consequently, the divergences can be removed
by the free field normal ordering.
 
Clearly, these arguments are perturbative in nature and they do not
change the fact that if some of the coefficients $\alpha_n$ with $n>2$
is non-zero ({\it i.e.}, if the theory is naively non-renormalizable) 
the derivative expansion breaks down in
the UV at the scale $M$ and the theory becomes strongly coupled at
that scale. Nevertheless, given that it is possible to renormalize the theory by
normal ordering, one may hope to get around this strong coupling and
extract some information non-perturbatively. It would be especially
interesting to consider the following family of 
Lorentz
invariant Lagrangians,
\be
\label{class}
S=\int dx^+ dx^-
\left[\d_+\psi\d_-\psi
+M^2U(M^{-1}\d_+\e^\psi)
\right]
\;
\ee
generalizing the Einstein-aether action (\ref{Spsi0}).

An immediate consequence of the fact that (\ref{Spsi0}) is
renormalized by normal ordering is that the coupling $g^2$ remains constant
to all
orders in the perturbation theory
and the parameter $\beta$ is renormalized multiplicatively. Given that
any particular value of $\beta$ has no physical meaning and can be changed
by the shift of $\psi$ we conclude that the classical scale symmetry
(which corresponds to the choice $f=\alpha x^+$, $g=\alpha x^-$ 
in (\ref{conf})) is preserved also at the
quantum level. 

The arguments about renormalizability by normal ordering presented in
this section can be easily generalized to the case of Lagrangians
containing several fields. In particular, they are applicable to the
model (\ref{Spsichi}). Besides, it is straightforward to check that
the model (\ref{Spsichi}) at the classical level enjoys the
``half-conformal'' symmetry (\ref{conf}) with the field $\chi$
transforming as a usual scalar boson. Therefore, the model (\ref{Spsichi})
provides an example of  a theory that obeys the $x^+$-ordered causal
structure, is classically invariant under the symmetry (\ref{conf})
and can be renormalized by normal ordering. These properties suggest that the model (possibly, with
some modifications) may be exactly solvable.  

\subsection{Case $\beta_{(+)}=\beta_{(-)}=\beta>0$}
\label{sec:equalbeta}

This case is parity symmetric and consequently it is not directly
related to the main subject of the present paper, {\it i.e.} the
investigation of the $x^+$-ordered causal structure that is
incompatible with parity. Still, for
completeness, we discuss this case as well. 

\begin{figure}[t]
\centering
\includegraphics[viewport=00 220 600 590,width=0.4\textwidth,clip]{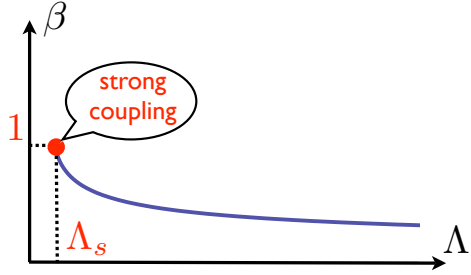}
\caption{\small\it One-loop running for $\beta_{(+)}=\beta_{(-)}$.}
\label{fig:betarunning}
\end{figure}

Let us start with the analysis of the RG equation (\ref{RGE}) in this
case. As already mentioned, $\beta$ runs to zero in the UV.
In the IR it approaches unity at
some scale $\Lambda_c$ (see Fig.~\ref{fig:betarunning}). 
Naively,
Eq.~(\ref{cpsi}) 
implies that the theory becomes unstable when one goes to even longer
distance scale, where $\beta>1$. What actually happens is that the model
becomes strongly coupled at the scale $\Lambda_c$. To
see the origin of the strong coupling, note that at the scale
$\Lambda_c$ where $\beta$ approaches unity, the $\pi\pi$ propagator
in the background $\psi_0$
degenerates
\[
\langle\pi\pi\rangle=\frac{ig^2}{(p_++p_-)^2}=\frac{ig^2}{2p_0^2}\;.
\]
This implies that the propagation
velocity of the $\pi$ modes turns zero at the scale $\Lambda_c$. Equivalently, the gradient
term $(\d_x\pi)^2$ drops out from the quadratic action for
$\pi$. Consequently, one is forced to take into account the
higher dimensional operators in the effective action, such as 
$(\d^2_x\pi)^2$, and the perturbation theory breaks down.

Interestingly, this behavior is typical also for Lorentz breaking
effective theories in four
dimensions~\cite{ArkaniHamed:2003uy,Dubovsky:2004sg}. There is an
important difference, however. In four dimensions, inclusion of the
leading higher derivative contributions to the quadratic action is
sufficient to get around the breakdown of the perturbation theory. As a
result one obtains a low energy effective theory of weakly interacting
Goldstone bosons, albeit with an exotic dispersion relation
$\omega^2\propto k^4$.  This
theory is healthy, provided
the sign in front of the $(\d^2_x\pi)^2$ operator in the action is
negative. 

To see what happens instead in two dimensions, let us recall that the
modified dispersion relation changes the power counting for the
scaling dimensions of the operators in the effective theory
\cite{Polchinski:1992ed,ArkaniHamed:2003uy}. In the
case $\omega^2\propto k^4$, if one takes the scaling
dimension of energy to be equal to 1, the scaling dimension of the
spatial momentum is equal to $1/2$. It is straightforward to check
that the scaling dimension of the field $\psi$ becomes then 
$(D-3)/4$,
where $D$ is the space-time dimensionality. It is negative at $D=2$ so
that the effective field theory acquires an infinite number of relevant
operators (a simple example being $(\d_x\psi)^4$ coming from
$(\d_+\psi\d_-\psi)^2$). Consequently, the derivative expansion is
totally doomed and the theory is strongly coupled at $\Lambda_c$.

This behavior of the model --- the RG running from asymptotic freedom in
UV to strong coupling at a finite scale in IR --- is similar to that
of other two-dimensional sigma models, such as, e.g., the
$O(N)$ model \cite{Polyakov:1975rr}. Following this analogy
it is natural to expect that the strong coupling leads to generation
of a mass gap
$\sim\Lambda_c$ and restoration of the Lorentz symmetry at this scale.
On the other hand, the
physics at short distances can be accurately described by a
perturbation theory around the symmetry breaking semiclassical vacuum
$\psi=0$ (or $\psi=const$---all these vacua are related by boosts).

\begin{figure}[t]
\centering
\includegraphics[width=0.45\textwidth,clip]{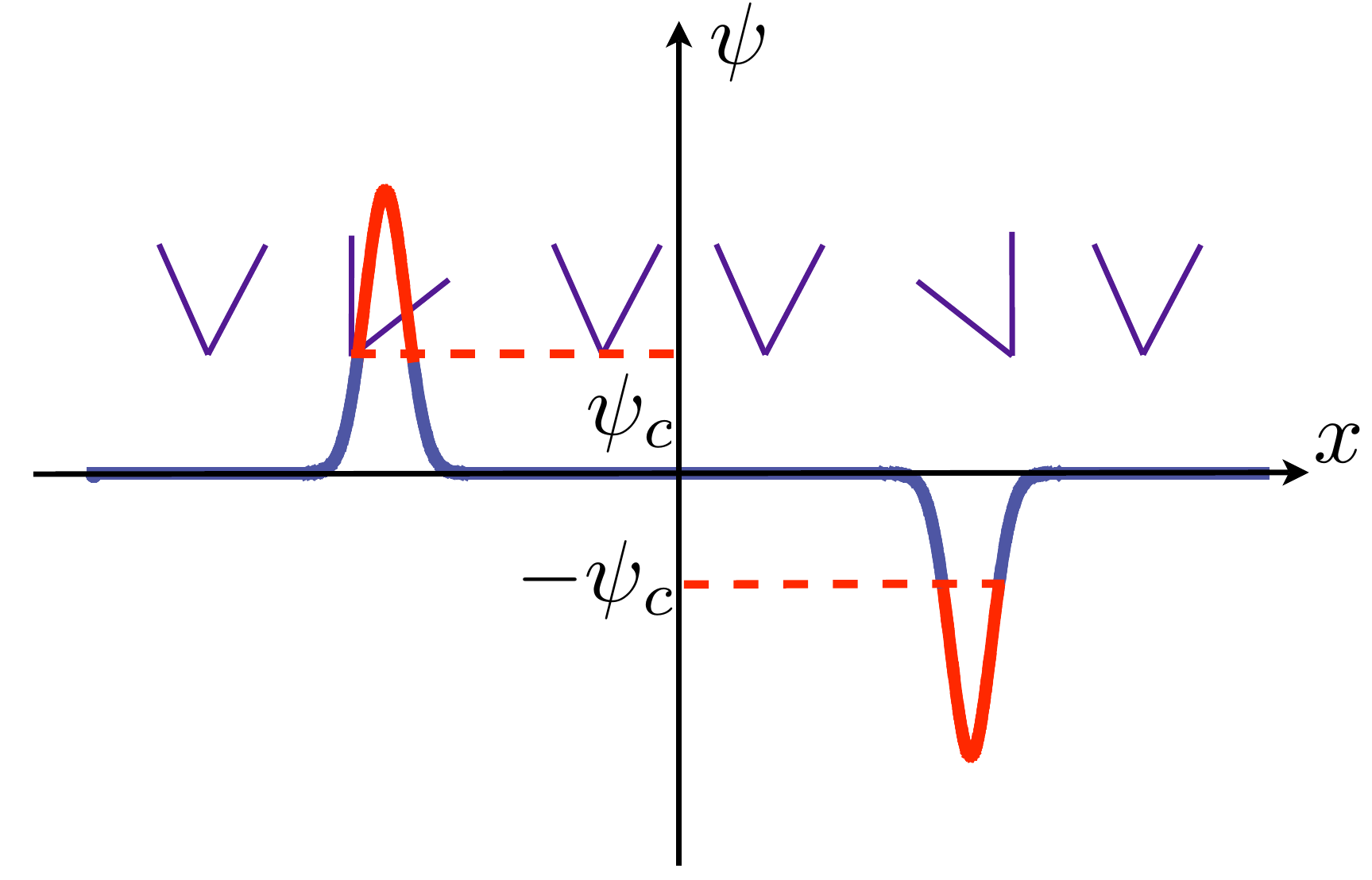}
\caption{\small\it An example of a static field configuration with zero energy and momentum. Spikes with $|\psi|>\psi_c$
have negative energy density that cancels  positive energy outside the spikes. The total momentum is zero because the field configuration is odd under $x\to-x$ parity. The causal cones of the effective metric for 
$\psi$ perturbations are shown schematically. Each spike acts as an event horizon 
for $\psi$ perturbations.}
\label{fig:spike}
\end{figure}

However, this physical picture should be taken with a grain of
salt. The reason is that, as mentioned in Sec.~\ref{sec:aether}, the
light-cone Hamiltonian of the system is unbounded from below which may
indicate a non-perturbative instability. Let us investigate this problem
using the conventional coordinates $(t,x)$. 
The Noether
procedure gives rise to the following expressions for the conserved
energy $H$ and momentum $P$ of the $\psi$ field,
\begin{align}
\label{hamiltonian}
&H=\frac{1}{2g^2}\int dx\left\{\l1+\beta\cosh{2\psi}\r\l\d_t\psi\r^2+
\l1-\beta\cosh{2\psi}\r\l\d_x\psi\r^2\right\}\;,\\
\label{momentum}
&P=\frac{1}{g^2}\int dx\left\{\l1+\beta\cosh{2\psi}\r\d_t\psi\d_x\psi
+\beta\sinh{2\psi}\l\d_x\psi\r^2\right\}\;.
\end{align}
Note that the gradient
energy becomes negative for large field values suggesting that the
semiclassical vacuum is indeed unstable at the non-perturbative
level. Let us see what are the field configurations the vacuum can
decay to, namely let us construct sample field configurations that
have a form of localized excitations around $\psi=0$ vacuum at $t=0$
and carry zero energy and momentum. One can set the time derivative of the field to
zero at $t=0$, 
\be
\label{nodt}
\d_t\psi(0,x)=0\;,
\ee
so that the kinetic energy and the first term in the expression
(\ref{momentum}) for the spatial momentum vanish. Then a simple
example of a localized field configuration with zero energy is a high
enough spike, such that the negative gradient energy in the region
with $\psi>\psi_c$, where
\be
\label{psicr}
\psi_{c}=\frac{1}{2}\cosh^{-1}{(1/\beta)}\;,
\ee
 compensates for the positive gradient energy in the
region with $\psi<\psi_c$ (see Fig.~\ref{fig:spike}). 
 Note that in
principle a spike need not be narrow and may be more like a bump; all what one needs is a region with
$\psi>\psi_c$, such that the spatial field gradient does not vanish
everywhere inside this region and compensates for the positive gradient
energy outside. Finally, note that it is impossible for a single spike
(or bump) to be spontaneously created from the vacuum as it carries a
non-zero spatial momentum, coming from the second term in
(\ref{momentum}). However, it is straightforward to make the total
momentum vanish by introducing the second spike, where the field
$\psi$ takes large {\it negative} values, $\psi<-\psi_c$. For instance, any field configuration
satisfying (\ref{nodt}) and odd with respect to spatial reflections,
$\psi(-x)=-\psi(x)$, has vanishing spatial momentum.

To understand better the physical interpretation of these spikes, let
us study the structure of the causal cones of the effective metric
(\ref{Gpm}) describing propagation of the scalar fluctuations in the
presence of a spike. At $\psi=0$ this is just a symmetric cone of a
subluminal particle that propagates with velocity (\ref{cpsi}) in both
directions. The transformation rule (\ref{psitrans}) implies that
taking $\psi>0 $ is equivalent to looking at the same situation from
the point of view of an observer who moves left with the rapidity
equal to $\psi$. Consequently, at $\psi>0$ the causal cone of the
$\psi$ particles is bended in the right direction. At $\psi>\psi_c$
this bending becomes so strong that both ``left'' and ``right'' moving
waves start propagating in the right direction (see
Fig.~\ref{fig:spike}). Consequently, the negative (or zero) energy
spike acts as a horizon for the $\psi$ signals.

It will be interesting to calculate the rate of the instability
towards producing field configurations of the type shown in Fig.~\ref{fig:spike} out of the vacuum. Presumably, at
small couplings this decay is described by a semiclassical bounce
solution, however, we have not been able to find this solution so
far. Still, assuming that such a solution exists, let us speculate on
possible implications of the vacuum decay process. Due to classical
scale invariance of the Einstein-aether action (\ref{Spsi}), 
the bounce solution is
not unique, rather there is a whole continuous family of solutions of
a different size $R$. On  dimensional grounds the vacuum decay rate has the form
\[
\Gamma\sim \int dR\, R^{-3}\e^{-S_{b}(R)}\;,
\]
where the dependence of the bounce action $S_{b}$ on $R$ arises
through the RG running of the couplings. The theory is free in the UV, so that
$S_{b}\to \infty$ as $R\to 0$. However, because of the $R^{-3}$
prefactor, it still may happen that the total decay rate is UV
dominated and diverges. This would
imply an inconsistency of the model. A more interesting option is
that $\Gamma$ is saturated at large values of $R$. Then, creation of
the field configurations as in Fig.~\ref{fig:spike} may provide a
natural mechanism for restoration of the Lorentz symmetry in the IR. 
Still, it is not clear if the system ends up in a well-defined vacuum
state or exhibits a runaway behavior.

To summarize, our analysis of the parity preserving case,
$\beta_{(+)}=\beta_{(-)}>0$, is somewhat inconclusive. This model can either
be plagued by non-perturbative instabilities or can be a consistent
quantum field theory analogous to the conventional two-dimensional
sigma-models with strong coupling and gap generation in the IR.

\subsection{Case $\beta_{(+)}=\beta_{(-)}\equiv\beta<0$}
\label{crazysuperluminal}
Naively one may expect that this model suffers from a non-perturbative instability similar to that present in
the $\beta_{(+)}=\beta_{(-)}>0$ case due to the fact that the coefficient in front of the $(\d_t\psi)^2$ term in the Hamiltonian 
(\ref{hamiltonian}) is not positive definite.  However, the above arguments
cannot be generalized to the present case. Indeed, it is the coefficient in front of the kinetic term in the Hamiltonian (\ref{hamiltonian}) that becomes negative now, so that it is impossible to find a configuration with negative energy
 while keeping $\d_t\psi=0$. More importantly, it is straightforward to check that when the field $\psi$ exceeds the critical value (\ref{psicr}), so that there is a chance to have negative energy, the causal cone of the effective metric (\ref{Gpm}) is bended back in time, so that the
$t=const$ surface is not space-like with respect to this metric any more and cannot be used to define the initial Cauchy data.

One may try to define a field configuration with vacuum quantum numbers on a
surface with a more complicated shape, such that it  is space-like with respect to the effective metric (\ref{Gpm}) even for $\psi>\psi_{c}$. However, it is straightforward to see that this is impossible.
Indeed,  if a slice has a normal vector 
\[
n_\alpha=(1,\theta)\;
\]
at some point,
then in order for this vector to be  timelike with respect to the effective metric (\ref{Gpm}) the following inequality should hold
\be
\label{spacelike}
G^{\alpha\beta}n_\alpha n_\beta>0\;,
\ee
where $G^{\alpha\beta}$ is the effective metric (\ref{Gpm}). In $(t,x)$ coordinates
it reads
\be
\label{Gtx}
G^{\alpha\beta}\d_\alpha\d_\beta=(1+\beta\cosh{2\psi})\d_t^2+2\beta\sinh{2\psi}\d_t\d_x+(-1+\beta\cosh{2\psi})\d_x^2\;.
\ee
On the other hand, the Noether energy density evaluated at this point can be presented in the following 
form
\be
\label{thetaE}
 T^\alpha_0n_\alpha={1\over 2g^2}\l G^{\alpha\beta}n_\alpha n_\beta\l\d_t\psi\r^2
 -G^{xx}(\d_x\psi-\theta\d_t\psi)^2\r\;,
\ee
where
\be
T^{\alpha}_\beta={1\over g^2}\l G^{\alpha\nu}\d_\nu\psi\d_\beta\psi-{1\over 2}\delta^\alpha_\beta
G^{\mu\nu}\d_\mu\psi\d_\nu\psi\r
\ee
 is the Noether energy-momentum tensor of the Einstein-aether model. Consequently, the energy density (\ref{thetaE}) is non-negatively definite on a spacelike  slice
as soon as $G^{xx}<0$, which is true for $\beta<0$. This suggests that at least at the semiclassical level 
the vacuum is stable.

These considerations appear encouraging.
Still there remains a question how to make sense of  a quantum  theory without a fixed causal structure.
Note that the same question arises in the case of quantum gravity\footnote{One can push the analogy between gravity and  the $\beta<0$ theory  even further: in both cases it is problematic
to define a consistent path integral due to the absence of  a positive definite Euclidean action.}.
 In our case the only possibility we see (apart from the option that the theory is pathological), is that
the spatial parity in spite of being unbroken at the level of the action is nevertheless broken spontaneously\footnote{Recall that the spontaneous breaking of a discrete symmetry is not forbidden by the CMW theorem.}. 
Indeed, this opens a room for the vacuum retarded Green's function to be Lorentz-invariant but parity violating. The two vacua related by spatial parity would be characterized
by the  $x^+$-ordered and  $x^-$-ordered retarded Green's functions respectively. 
Still the absence of the fixed causal structure would 
 manifest itself in the existence of 
more general  quantum states  that can be thought of as an admixture of finite size domains with different causal ordering. 

It definitely remains unclear whether it may be possible to make sense of such a situation. However, if the answer
is yes, this would be a prototype of a higher dimensional theory where superluminal propagation arises
due to spontaneous breaking of Lorentz symmetry as discussed in Sec.~\ref{subsec:generalities}. 
Spontaneous breaking of
 the continuous Lorentz symmetry is impossible in 2d, however, in this case it  is enough just for the spatial parity --- a discrete symmetry --- to be spontaneously broken.
 
 To conclude this discussion and  to avoid a possible confusion let us stress that  this
  controversial case is totally unrelated from our previous analysis
   of theories with fixed
 instantaneous causal structure. The consistency of the latter theories will not be affected by 
 the eventual conclusion on whether $\beta<0$ theory is pathological or not.

\section{Coupling to gravity}
\label{sec:gravity}
Let us discuss now what happens when the $SO(1,1)$ sigma-model
studied above is coupled to gravity. Such a coupling is
straightforward to introduce using the Einstein-aether form of the
action (\ref{EAction}).  One obtains 
\be
\label{Sgrav}
S_{gr}=-\frac{1}{2\pi\kappa}\int d^2x\sqrt{-g}\l R+\mu^2\r
+S_{EA}(g_{\mu\nu},V_\mu)\;,
\ee
where $S_{EA}(g_{\mu\nu},V_\mu)$ is the action (\ref{EAction})
minimally coupled to gravity. In the above expression $R$ is the Ricci
scalar and $\mu^2$ is the cosmological constant. In two dimensions the
curvature term in the gravitational action is purely topological, 
so at the classical level the metric
$g_{\mu\nu}$ acts as a Lagrange multiplier enforcing the constraint
$T_{\mu\nu}=0$. At the quantum level, in the conformal gauge
$g_{\mu\nu}=\e^{\phi}\eta_{\mu\nu}$, one obtains a conformal theory
described by the Liouville action for the dilaton $\phi$ coming from
the measure in the functional integral, plus the action for  matter (in the case
at hand $V_\mu$, $\lambda$)  ``gravitationally dressed'' by the
dilaton ~\cite{Polyakov:1981rd}--\cite{Ginsparg:1993is}. To see how
the Einstein--aether sector contributes to the dilaton action let us
have a closer look at the part of the functional integral involving
$V_\mu$ and $\lambda$. After fixing the conformal gauge it takes the
form 
\be
\label{EApath}
Z_{EA}=\int{\cal D}_\phi\lambda{\cal D}_\phi V_\mu\;
\e^{i\int d^2 x\left[\lambda\l 2V_+V_--\e^\phi\r+
\tilde{S}_{EA}(g_{\mu\nu}=\e^\phi\eta_{\mu\nu},V_\mu)\right]}\;.
\ee
where $\tilde{S}_{EA}(g_{\mu\nu},V_\mu)$ is the part of the action
independent of the Lagrange multiplier $\lambda$.  The
$\phi$-subscript in the integration measure indicates that it is defined
in such a way that (see, e.g., \cite{Alvarez:1982zi}),
\begin{subequations}
\label{measures}
\begin{align}
&\int{\cal D}_\phi\lambda
\e^{i\int d^2x\sqrt{|g|}\lambda^2}=
\int{\cal D}_\phi\lambda
\e^{i\int d^2x\,\e^\phi\lambda^2}=1\;,\\
&\int{\cal D}_\phi V_\mu\e^{i\int d^2x\sqrt{|g|} g^{\mu\nu}V_\mu V_\nu}
=\int{\cal D}_\phi V_\mu\e^{i\int d^2x\,\eta^{\mu\nu}V_\mu V_\nu}
=1\;.
\end{align}
\end{subequations}
Equations (\ref{measures}) imply that the measure for the vector
field $V_\mu$ is $\phi$-independent. On
the other hand, the
measure for the Lagrange 
multiplier $\lambda$ does depend on the dilaton $\phi$. This
dependence is the same as for an ordinary two-dimensional scalar
field. 
One gets rid of this
dependence at the expense of introducing the Jacobian factor
\[
\e^{iJ(\phi)}=\det{ \l\e^{\phi(x)}\delta(x-x')\r}\;,
\]
which contributes to the Liouville action for the dilaton
field ~\cite{Polyakov:1981rd}--\cite{Ginsparg:1993is}. Now it is
straightforward to perform integration over the Lagrange multiplier
$\lambda$ that results in $\delta(2V_+V_--\e^\phi)$. Finally, one 
integrates over $V_+$ and obtains
\be
\label{Aintegral}
\int {\cal D}V_-|\;\det{V_-^{-1}\delta(x-x')}|
\e^{
i\tilde{S}_{EA}
(\e^\phi\eta_{\mu\nu},V_-,V_+=\e^\phi V_-^{-1})
}
=
\int{\cal D}\psi\; \e^{
i\tilde{S}_{EA}
(\e^\phi\eta_{\mu\nu},V_-=\e^{\psi}/\sqrt{2},V_+=\e^{\phi-\psi}/\sqrt{2})
}\;.
\ee
Explicitly, the action in the last formula has the form
\be
\label{Sphipsi}
\tilde S_{EA}(\phi,\psi)=\frac{1}{g^2}\d_+\psi\d_-(\psi-\phi)
+\frac{\beta_{(+)}}{2g^2}\e^{2\psi-\phi}(\d_+\psi)^2
+\frac{\beta_{(-)}}{2g^2}\e^{\phi-2\psi}\l\d_-(\psi-\phi)\r^2\;.
\ee
To summarize, when coupled to gravity, in the conformal gauge the
Einstein-aether sector contributes to the Liouville action in the same
way as a single boson; additionally
there is an interaction between the $SO(1,1)$ Goldstone $\psi$ and the
dilaton due to the fact that the Einstein-aether action is not
conformally invariant. 

Note, that the dilaton does not receive a second order kinetic term with
respect to the light cone time $x^+$. 
 As
discussed in Sec.~\ref{sec:2d}, this indicates a subtlety with the quantization of the theory
(\ref{Sphipsi}) according to the $x^+$-ordered causal structure.
Namely, the canonical formalism for this theory involves first class constraints
related to the residual gauge  invariance  of the action  (\ref{Sphipsi}).
This gauge invariance manifests itself  in
the conformal symmetry, 
that at the classical level acts as
\begin{align}
&\phi(x^+,x^-)\mapsto\phi(f(x^+),g(x^-))+\log f'+\log g'\\
&\psi(x^+,x^-)\mapsto\psi(f(x^+),g(x^-))+\log g'\;.
\end{align}
Clearly, more work is needed to understand the physics of
(\ref{Sphipsi}). Note that the 
light cone gauge
\cite{Polyakov:1987zb,Knizhnik:1988ak} may turn out to be more appropriate than the
conformal one for the analysis of theories with instantaneous causal
structure coupled to gravity.

\section{Conclusions and future directions}
\label{sec:conclusions}
What are the main lessons to be drawn from our study? The parity breaking
theories described here, to the best of our
knowledge, present the first 
example of Lorentz invariant quantum field theories
exhibiting non-local physics. Thus, they provide a theoretical
laboratory for studying the effects of non-locality.
Admittedly, our results crucially rely on the peculiar properties of a
two dimensional world and there is no direct way to generalize them to
higher dimensions. However, studies of two dimensional field theories
proved to be useful for modeling dynamics in higher dimensions in many
cases, such as confinement in QCD~\cite{'t Hooft:1974hx} and black
hole information paradox~\cite{Strominger:1994tn}. Although at the
level of intermediate calculations two dimensional physics is rather
special in these cases as well, many of the final results turn out to
be universal. So one may hope that also in the case of a superluminal
travel two dimensional models may provide useful universally
applicable lessons.
In fact, the models considered in this paper share
many of the properties of effective field theories describing
spontaneous Lorentz breaking in higher dimensions
\cite{ArkaniHamed:2003uy,Dubovsky:2004sg,Jacobson:2000xp}.

Clearly, to proceed further in this direction the key question is
whether the instantaneous theories remain consistent when coupled to
gravity, and if the answer is yes, what the physical properties of
the resulting theories are. It would be especially interesting  to introduce
the instantaneous causal structure in the dilaton gravity models
possessing black hole solutions like those discussed
in~\cite{Strominger:1994tn,Grumiller:2002nm}. If this can be done, one may obtain an
example of a radical resolution of the black hole information paradox
by allowing the information to escape already at the classical level,
as it happens at the effective theory level in the Higgs
phases of gravity in four dimensions~\cite{Dubovsky:2007zi}.

There is another, even more intriguing, way to look at the
instantaneous two dimensional models.  Namely, in principle any model
of two dimensional gravity can be thought of as a world-sheet theory
for a (non-critical) string. In the present case one would obtain a
string theory with instantaneous
interactions propagating along the world-sheet. In such a
situation one expects to find strong non-localities from the target
space point of view as well, so this may provide a way to generate
non-local gravitational theories in higher dimensions. In fact, if one
thinks that effective field models describing gravity in the Higgs
phase can be UV completed, this is a natural direction to
proceed. Indeed, a natural way to obtain a weakly interacting quantum theory
of a massless spin-2 particle is to start with a string. 
To find a deformation of such a theory one modifies the
world-sheet dynamics. Usually, changing the world-sheet action
corresponds just to taking a different background of the underlying
string theory and such a modification is not radical enough to
reproduce the physics of
\cite{ArkaniHamed:2003uy,Dubovsky:2004sg,Jacobson:2000xp}. A drastic
change of the causal structure of the world-sheet, if it can be
implemented in a consistent manner, definitely appears as a more
radical deformation of a theory. As an example of a possible 
world-sheet action
one can take the action (\ref{Spsichi}) with several
flavors $X^i$ ($i=1,\dots D$) and couple it to gravity in the
conformal gauge as in (\ref{Sphipsi}); this yields
\be
\label{string}
S(\phi,\psi,X^i)\!=\!\!\int\! d^2x\left\{\frac{1}{g^2}\left[\d_
+\psi\d_-(\psi-\phi)-\d_+X^i\d_-X^i\right]
+\frac{\beta}{2g^2}\e^{2\psi-\phi}\left[(\d_+\psi)^2+(\d_+X^i)^2\right]
\right\}
+S_{L}(\phi)\;,
\ee
where $S_L(\phi)$ is the Liouville action. 

Both if one is interested in instantaneous theories
coupled to 2d gravity as toy models for non-local gravitational
theory, or as a world-sheet description of an exotic string theory,
the crucial step is to construct conformal field theories exhibiting the
instantaneous causal structure. At the classical level 
the action (\ref{string}) is
conformal, but it remains to be seen whether the conformal symmetry
survives at the
quantum level. Constructing 
a unitary conformal field theory with instantaneous
causal structure  would be interesting from
yet another point of view. Indeed, using $AdS_3/CFT_2$ duality such a
theory should be dual to a three-dimensional gravitational theory with
instantaneous signal propagation.  This proposal is similar to that of
\cite{Sundrum:2007xm}, where it was also suggested to introduce
instantaneous interaction on the CFT side as a way of generating
non-local gravitational theories.  The important difference is that in
the current context this can, in principle, be achieved without breaking
Poincare invariance.

To summarize, two-dimensional models with instantaneous Poincare
invariant causal structure provide interesting examples of 
quantum field theories with non-local behavior. They provide a rich framework to study
non-local physics, even in the presence of gravity.  They suggest a
number of intriguing opportunities, and we briefly outlined some of
them. We feel that it is worth pursuing these ideas: irrespectively 
of whether they turn out to be successful or not, from this exercise
one will learn more about both quantum field theory and gravity.

\subsection*{Acknowledgements} It is a pleasure to
thank Nima Arkani-Hamed, Chris Beasley, Fedor Bezrukov, Eric D'Hoker, 
Jacques Distler,
Gia Dvali, Willy Fischler, Ted Jacobson, Randy Kamien, Oleg Lebedev,
Alex Maloney, David Nelson, Take Okui, Sasha Penin,
Riccardo Rattazzi,  Michele Redi, Valery Rubakov, Andy Strominger
and Giovanni Villadoro for useful discussions. The work of S.S. was
supported in part by EU 6th Framework Marie Curie
Research and Training network ``UniverseNet'' (MRTN-CT-2006-035863). This work was also partially supported
by the NSF Grant No. PHY-0455649.


\end{document}